\documentclass[12pt]{article}

\newcommand{\beq}{\begin{equation}}
\newcommand{\eeq}{\end{equation}}
\newcommand{\beqa}{\begin{eqnarray}}
\newcommand{\eeqa}{\end{eqnarray}}
\newcommand{\beqar}{\begin{eqnarray*}}
\newcommand{\eeqar}{\end{eqnarray*}}

\newcommand{\al}{\alpha}
\newcommand{\be}{\beta}

\def\spa          {\ \ \ }
\def\non          {\nonumber}
\def\ha           {\mbox{$\frac{1}{2}$}}

\def\spa          {\ \ \ }
\def\mand         {\spa\mbox{and}\spa}

\def\Tr           {\mbox{\rm Tr}\,}

\def\cd           {{\cdot}}
\def\ran          {\rangle}
\def\lan          {\langle}
\def\fsH    {H\!\!\!\!/\,}

\newcommand{\eps}{\epsilon}
\newcommand{\ga}{\gamma}

\newcommand{\inn}{\!\cdot\!}

\newcommand{\lam}{\lambda}

\newcommand{\z}{\zeta}

\newcommand{\labell}[1]{\label{#1}} 
\newcommand{\reef}[1]{(\ref{#1})}
\newcommand\prt{\partial}

\newcommand\cL{{\cal L}}
\newcommand\cD{{\cal D}}

\newcommand\bz{\bar{z}}

\newcommand\cT{T}

\begin{document}
\baselineskip 16.8pt%
\begin{titlepage}
\vspace*{1mm}%
\hfill%
  \halign{#\hfil         \cr
         CERN-PH-TH/2012-341\cr
           } 
\vspace*{11mm}%
\center{ {\bf \Large On D-brane anti D-brane effective actions and their corrections to all orders in alpha-prime
}}\vspace*{.7mm}
\begin{center}
{Ehsan Hatefi }$\footnote{E-mail:ehatefi@ictp.it,ehsan.hatefi@cern.ch}$

\vspace*{0.3cm}{ {\it
International Centre for Theoretical Physics\\
 Strada Costiera 11, Trieste, Italy
  \\
  and

  Theory Group, Physics Department, CERN\\
 CH-1211, Geneva 23, Switzerland
}}
\vspace*{.72cm}
\end{center}
\begin{center}{\bf Abstract}\end{center}
\begin{quote}
Based on a four point function, the S-matrix elements at disk level of the scattering amplitude of one closed string Ramond-Ramond field ($C$), two tachyons and one scalar field, we find out new couplings in brane anti brane effective actions for $p=n,p+2=n$ cases. Using the infinite corrections of the vertex of one RR, one gauge and one scalar field and applying the correct expansion, it is investigated in detail how  we produce the infinite gauge poles of the amplitude for $p = n$ case. By discovering new higher derivative corrections of  two tachyon-two scalar couplings in brane anti brane systems  to all orders in $\alpha'$, we also obtain the infinite scalar poles in $(t'+s'+u)$-channel in field theory. Working with the complete form of the amplitude with the closed form of the expansion and comparing all the infinite contact terms of this amplitude, we  derive several new Wess-Zumino couplings with all their infinite higher derivative corrections in the world volume of  brane anti brane systems. In particular, in producing all the infinite  scalar poles of  $<V_C V_{\phi} V_T V_T>$,  one has to consider the fact that scalar 's vertex operator in $(-1)$-picture  must carry the internal $\sigma_3$ Chan-Paton matrix. The symmetric trace effective action has a non-zero coupling between $D\phi^{(1)i}$ and $D\phi^{(2)}_i$ while this coupling does not exist in ordinary trace effective action.

\end{quote}
\end{titlepage}
\section{Introduction}

The relation between closed string Ramond-Ramond field and D-branes was widely realized in \cite{Polchinski:1995mt}. It is worthwhile to refer some of the main references on branes \cite{Witten:1995im,Polchinski:1996na,Bachas:1998rg}.

\vskip .1in

Concerning T-duality transformation, two important technical facts in the introduction of \cite{Hatefi:2012zh} have been addressed. However for good reasons we re-express the fact that for mix open-closed amplitudes (including closed string Ramond-Ramond) not all the coefficients of higher derivative corrections can be fixed in a correct manner, thus the direct calculations  will be necessary ( as we clarify in detail this argument in section 3.3 of this paper).

\vskip .1in

To review within full details Dirac-Born-Infeld and Wess-Zumino effective actions  \cite{Hatefi:2010ik,Myers:1999ps,Li:1995pq,Douglas:1995bn,Green:1996dd}  and all their references  might be considered.

\vskip .2in

In this paper we would like to deal with non-BPS branes and in particular we look for new couplings (which can be explored just by S-matrix calculations) in the world volume of brane anti brane systems in type II super string theory, where spatial dimension of $D_{p}$-branes becomes odd (even) for IIA (IIB) string theory.

\vskip .1in

Studying tachyons in super string theories may provide some good information about these theories in some backgrounds which are time-dependent \cite{Gutperle:2002ai,Sen:2002nu,Sen:2002in,Lambert:2003zr,Sen:2004nf}. Based on some arguments \cite{Sen:2004nf}, we highlight the point that the effective theory of all non-BPS branes does involve just massless states and tachyon. To be able to proceed with non-BPS branes in string theory within their applications, one should take into account \cite{Sen:2002nu,Sen:2002in,Sen:1999mg}.  The complete form of the effective actions of non-BPS branes was introduced in \cite{Garousi:2008ge,Hatefi:2012wj}.

\vskip .2in

 The only consistent effective action for D-brane anti D-brane systems, based on direct S-matrix computations of one closed string Ramond-Ramond, two tachyons and one gauge field  was appeared in \cite{Garousi:2007fk}. In that paper we have shown that there was a non-zero coupling between $F^{(1)}$ and $F^{(2)}$  and we found all their infinite higher derivative corrections.

\vskip .2in

 Note that for ordinary trace prescription, this coupling does not exist. It is important to emphasize that the effective action in \cite{Sen:2003tm} is not consistent with S-matrix computations and in fact symmetrized trace works out for super string computations. The reasons for such an argument ( see \cite{Garousi:2007fk}) are as follows.
 By using the effective action appeared in \cite{Sen:2003tm}, we are not able to produce all the tachyon poles of the amplitude of one RR, two tachyons and one gauge field, in addition to that if we use \cite{Sen:2003tm}, the structures and  some of the forms of  new couplings  (such as $F^{1}.F^{2}$ which is confirmed by S-matrix in \cite{Garousi:2007fk}) will be disappeared.

 \vskip .2in

 In this paper we explore the presence of  new term involving
 \beqa
 D\phi^{i(1)}.D\phi^{(2)}_i
  \nonumber\eeqa

  in DBI action of brane anti brane systems where this new coupling and its all order $\alpha'$ higher derivative corrections can be discovered just by applying S-matrix method of this paper.  Having set the tachyon to zero, both ordinary and symmetric trace effective actions become equivalent.

\vskip .2in

  In order to observe the details on tachyon condensation for brane anti brane system \cite{Sen:1998sm} has been constructed, moreover some of the the Ramond-Ramond couplings on the world volume of brane anti brane have been discussed within all needed details in \cite{Kennedy:1999nn,Garousi:2007fk}. Although several motivations including their explanations have been explained in \cite{Hatefi:2012wj}, however, it is important to emphasize some of them very briefly and for the rest of motivations we just refer to all references that appeared in \cite{Hatefi:2012wj}.

\vskip .1in

  The first motivation is actually related to dualities \cite{Sen:1998rg}. The second one somehow should be taken as a matter of brane production in which we are not going to go through details so just to clarify more \cite{Bergman:1998xv,Witten:1998cd} might be useful to look in. Indeed if the distance between brane anti brane is less than string's scale, then we will have two real tachyons in brane anti brane 's spectrum. The third motivation is to have inflation in string theory, one might consider brane anti brane effective action  \cite{Garousi:2007fk} in which brane and anti brane have to be separated \cite{Dvali:1998pa,Burgess:2001fx,Choudhury:2003vr,Kachru:2003sx},
 in which this formalism  was explained in detail in the section 2.1 of  \cite{Hatefi:2012wj}. The fourth motivation for studying tachyons is to work out with holographic models in QCD \cite{Casero:2007ae,Dhar:2007bz}. The last motivation is related to scattering amplitude method. Indeed the  S-Matrix computation does have a very strong potential to explore several new couplings with having no on-shell ambiguity.

\vskip .1in

  Having set some arguments \cite{Kraus:2000nj,Kennedy:1999nn}, one can talk about the Wess-Zumino effective action for the tachyons by applying super connection approach \cite{Roepstorff:1998vh}.

  Note that the super connection's curvature just for brane anti brane has been achieved in \cite{Garousi:2007fk} and it is generalized in \cite{Hatefi:2012wj} to actually take various couplings between gauge, tachyon and Ramond-Ramond field (RR) in the branes' world volume directions.

\vskip .1in

However the couplings between scalar field, RR and tachyons can not be derived by making use of the multiplication rule of the super matrices. The only way to discover these new couplings on the longitudinal and transverse directions of the brane anti brane is just based on scattering amplitude techniques. In \cite{Hatefi:2012zh,Hatefi:2010ik,Hatefi:2012ve} we found all the infinite corrections to BPS branes and in particular, special attention was paid to a conjecture on higher derivative corrections for both BPS and non-BPS branes in \cite{Hatefi:2012rx} where their applications to some dualities \cite{Hatefi:2012bp}, Ads/CFT correspondence and in M-theory
\cite{Hatefi:2012sy} were argued. Moreover in \cite{deAlwis:2013gka} we have analyzed in a type IIB flux compactification the transitions between two different vacua via D5/NS5 brane nucleation, more significantly some comparisons with KKLT and type IIA have been made.

\vskip .2in

In the next section we would like to compute in detail the  amplitude of one RR, two tachyons and one scalar field in the world volume of brane anti brane systems. Then we move on to momentum expansion and discuss about a particular expansion.

 \vskip .2in

 We do have several goals for carrying out this computation. One of the important goals of this paper is to explicitly find out (by comparing the S-Matrix of $C\phi TT$ with field theory analysis) the presence of new terms including
 \beqa
 D\phi^{i(1)}.D\phi^{(2)}_i
 \nonumber\eeqa

  in DBI action of brane anti brane systems where  $\phi^{(1)}$ lives on brane and $\phi^{(2)}$  moves on anti brane or viceversa.

  \vskip .1in

 The second goal will be  deriving (with explicit computations) the effective actions of brane anti brane systems and also  discovering several new Wess-Zumino couplings with all their infinite $\alpha'$ corrections. In section 3.1 we find all the infinite u-channel gauge poles with all infinite contact interactions for $p=n$ case in field theory. We find some new couplings like
 \beqa
 \partial_{i} C_{p-1} \wedge DT \wedge DT^* (\phi^1+\phi^2)^i
 \nonumber\eeqa
  we also obtain its infinite higher derivative corrections and fix its coefficient by making use of the $C \phi TT$ amplitude.
  \vskip .1in
 The third goal is as follows.
 We need to go on to
achieve all infinite  $\alpha'$ -higher derivative corrections of two tachyon, two scalar couplings  of brane anti brane systems and in particular we clearly show that these new corrections of brane anti brane systems  are completely different from non-BPS branes \cite{Hatefi:2012wj}, more importantly  their coefficients can not be read from higher derivative corrections of two tachyons and two gauge fields of brane anti brane systems \cite{Garousi:2007fk} (detailed  arguments will be highlighted in section 3.3 ). We explore these new two tachyon, two scalar couplings  of brane anti brane systems and clearly check  them by producing all infinite scalar poles of our string amplitude in $(u+s'+t')$- channel  for $p+2=n$ case.
For $p+2=n$ case we also obtain  a new coupling like
\beqa
\epsilon^{a_0 \cdots a_p} H_{ia_0 \cdots a_p} (\phi^{(1)}-\phi^{(2)})^i TT^*
\nonumber\eeqa
 and fix its coefficient as well.

 \section{The four point amplitude between one RR, one scalar field and two tachyons $(C\phi
TT)$ }

  In order to find the infinite higher derivative corrections of  two tachyon, two scalar couplings in the world volume of brane anti brane systems with exact coefficients, we have to have the complete form of the amplitude of $C \phi TT$. One might wonder whether  two scalar two tachyon couplings of non-BPS branes \cite{Hatefi:2012wj} could be applied to  brane anti brane systems. With in detail we will show that those corrections of non-BPS branes  give rise inconsistent results at each order of $\alpha'$ and in fact by using \cite{Hatefi:2012wj} we are not able to  actually  match neither first simple pole nor infinite scalar poles of the string amplitude of $C \phi TT$ in the world volume of brane anti brane systems with its field theory analysis.

  \vskip .1in

  In this paper within details we will point out that two scalar two tachyon couplings of brane anti brane systems to all orders are completely different from two tachyon two scalar couplings of non-BPS branes \cite{Hatefi:2012wj} and indeed we observe in a clear way that  making use of those non-BPS couplings we get wrong result and we can not match all infinite scalar poles of the string theory amplitude.

  \vskip .1in

  With this remarkable motivation we start discovering in detail the S-matrix elements of $C \phi TT$.

  \vskip .2in

The four point amplitude between one closed string RR, one gauge field and two tachyons $CATT$ in detail in \cite{Garousi:2007fk} has been done.  In this paper we show that apparently one can read the general structure of higher derivative corrections of two scalars two tachyons from  two tachyons and two gauge fields but there is a very important fact about corrections as follows.
  \vskip .1in
 All the coefficients of higher derivative corrections (just for mixed open-closed amplitude) can just be fixed by S-matrix method and not by any other method like T-duality transformation to the old computations (see footnote 1 of \cite{Hatefi:2012ve}) \footnote{We thank J.Polchinski and R.C.Myers for various discussions on this subject. }. Setting all physical state conditions we observe that there
are various subtleties in the mixed open-closed tree level amplitudes where we will point out some of them later on.
\vskip .2in
Nevertheless, let us  make some concrete arguments based on field theory analysis.
Therefore before getting to the computations, we want to
make some comments about apparent similarities between $CATT$ and $C \phi TT$ amplitudes. The amplitude
for $C\phi TT $  for $p=n$ case does include the infinite gauge poles and also for $p+2=n$
case, this amplitude does involve the infinite  scalar
poles. Notice that due to some kinematic constraints $C \phi TT$ does not have
any tachyon pole while  $CATT$ in addition to infinite gauge poles has infinite tachyon poles for $p=n$ case.

The other fact which comes out from the direct computation is that for $C \phi TT$ amplitude neither in $p=n$ nor
 in $p+2=n$ case we do not have any double pole.
 This is important to highlight it such that after some computations we found a double pole for  $p=n$ case in
 $CATT$ amplitude.
 The other fact is that the S-matrix of $C \phi TT$ will make sense for $p=n,p+2=n$ cases
while the amplitude of $CA TT$ was made of $p-2=n,p=n$ cases.

  \vskip .2in

The last motivation for performing the calculations of $C \phi TT$  is as follows.
 \vskip .1in

We could discover several new Wess-Zumino couplings meanwhile they are completely absent in $CATT$
amplitude \cite{Garousi:2007fk}. Therefore by some well known Conformal Field Theory methods in the
world volume of brane anti brane systems ( in type II super string
theory), we start finding
S-matrix of  two tachyons, one scalar field and one closed string RR field to indeed derive
 new couplings with all their higher derivative corrections to all orders of $\alpha'$ where the idea of obtaining closed form of higher derivative corrections to all orders  of $\alpha'$ sounds interesting.

 \vskip .2in

This $C \phi TT$ amplitude in brane anti brane system's world volume
can be written down as follows

\beqa
{\cal A}^{C \phi TT} & \sim & \int dx_{1}dx_{2}dx_{3}dzd\bar{z}\,
  \lan V_{\phi}^{(0)}{(x_{1})}
V_{T}^{(0)}{(x_{2})}V_T^{(0)}{(x_{3})}
V_{RR}^{(-\frac{3}{2},-\frac{1}{2})}(z,\bar{z})\ran.
\nonumber\eeqa

Note that we do have two more freedoms to find out the amplitude of $C \phi TT$ , like  working out either with
\beqa
\lan V_{\phi}^{(-1)}{(x_{1})}
V_{T}^{(0)}{(x_{2})}V_T^{(0)}{(x_{3})}
V_{RR}^{(-\frac{1}{2},-\frac{1}{2})}(z,\bar{z})\ran
\nonumber\eeqa
or

 \beqa
 \lan V_{\phi}^{(0)}{(x_{1})}
V_{T}^{(-1)}{(x_{2})}V_T^{(0)}{(x_{3})}
V_{RR}^{(-\frac{1}{2},-\frac{1}{2})}(z,\bar{z})\ran
\nonumber\eeqa

One remarkable fact about  tachyon's vertex operator in string theory is that it does relate to two real  components of a complex tachyon in field theory which means that the following relation holds
\beqa
T&=&\frac{1}{\sqrt{2}}(T_1+iT_2)\eeqa

In order to avoid some details we try to look for $C \phi TT$ amplitude in the following picture
\beqa
{\cal A}^{C\phi TT} & \sim & \int dx_{1}dx_{2}dx_{3}dzd\bar{z}\,
  \lan V_{\phi}^{(-1)}{(x_{1})}
V_{T}^{(0)}{(x_{2})}V_T^{(0)}{(x_{3})}
V_{RR}^{(-\frac{1}{2},-\frac{1}{2})}(z,\bar{z})\ran,\labell{sstring}\eeqa

Indeed it is the fastest way to carry out $C \phi TT$ amplitude  without making use of applying several physical state conditions and Bianchi identities. To clarify all things once more in this paper we express the general form of vertices for our amplitude\footnote{In
string theory, one does set $\alpha'=2$.}
\beqa
V_{T}^{(0)}(x) &=& \alpha'ik.\psi(x)  e^{\alpha'ik\cd X(x)}\lam\otimes\sigma_1\labell{vertex1}\\
V_\phi^{(-1)}(y)&=&\xi_i e^{-\phi(y)}\psi^i(y)e^{\alpha'iq\inn X(y)}\lam\otimes\sigma_3\nonumber\\
V_{RR}^{(-1)}(z,\bar{z})&=&(P_{-}\fsH_{(n)}M_p)^{\al\be}e^{-\phi(z)/2} S_{\al}(z)e^{i\frac{\alpha'}{2}p\cd X(z)}e^{-\phi(\bar{z})/2} S_{\be}(\bar{z}) e^{i\frac{\alpha'}{2}p\cd D \cd X(\bar{z})}\otimes\sigma_3\nonumber
\eeqa

Note that due to non-zero couplings between two tachyons and one RR in brane anti brane systems \cite{Garousi:2007fk,Kennedy:1999nn}  the Chan-Paton (CP) factor for RR for brane anti brane in  (-1)-picture (which is
$\sigma_3 $) is different from CP factor for RR in non-BPS branes which is $\sigma_3\sigma_1$.

 \vskip .1in

 $k$ is tachyon's  momentum which satisfies
$k^2=\frac{1}{4}$ and physical state conditions for scalar is $k_2.\xi=k_1.\xi=q.\xi=0$. The scalar in (-1)-picture does accompany  $\sigma_3$ factor, the so called  internal degree of freedom.
Notice that the other CP factors of  open strings and RR closed string for diverse pictures are  introduced in \cite {Hatefi:2012wj} . Thus our amplitude has   $\Tr(\sigma_3\sigma_3\sigma_1\sigma_1)=2$ factor.
The projector in closed string is $P_{-} = \ha (1-\ga^{11})$  and
\begin{displaymath}
\fsH_{(n)} = \frac{a
_n}{n!}H_{\mu_{1}\ldots\mu_{n}}\ga^{\mu_{1}}\ldots
\ga^{\mu_{n}}
\ ,
\non\end{displaymath}

with $n=2,4,a_n=i$ ($n=1,3,5,a_n=1 $) for  IIA (IIB).
 Having used doubling trick  \cite{Hatefi:2012wj}, we now replace  fields to a total complex plane which means that the following change of variables have to be considered
\begin{displaymath}
\tilde{X}^{\mu}(\bar{z}) \rightarrow D^{\mu}_{\nu}X^{\nu}(\bar{z}) \ ,
\spa
\tilde{\psi}^{\mu}(\bar{z}) \rightarrow
D^{\mu}_{\nu}\psi^{\nu}(\bar{z}) \ ,
\spa
\tilde{\phi}(\bar{z}) \rightarrow \phi(\bar{z})\,, \mand
\tilde{S}_{\al}(\bar{z}) \rightarrow M_{\al}{}^{\be}{S}_{\be}(\bar{z})
 \ ,
\non\end{displaymath}

 with
\begin{displaymath}
D = \left( \begin{array}{cc}
-1_{9-p} & 0 \\
0 & 1_{p+1}
\end{array}
\right) \ ,\,\, \mand
M_p = \left\{\begin{array}{cc}\frac{\pm i}{(p+1)!}\ga^{i_{1}}\ga^{i_{2}}\ldots \ga^{i_{p+1}}
\eps_{i_{1}\ldots i_{p+1}}\,\,\,\,{\rm for\, p \,even}\\ \frac{\pm 1}{(p+1)!}\ga^{i_{1}}\ga^{i_{2}}\ldots \ga^{i_{p+1}}\ga_{11}
\eps_{i_{1}\ldots i_{p+1}} \,\,\,\,{\rm for\, p \,odd}\end{array}\right.
\non\end{displaymath}

Once more, it is worth pointing out the needed correlators  $X^{\mu},\psi^{\nu}, \phi$, as below
\begin{eqnarray}
\lan X^{\mu}(z)X^{\nu}(w)\ran & = & -\frac{\alpha'}{2}\eta^{\mu\nu}\log(z-w) \ , \non \\
\lan \psi^{\mu}(z)\psi^{\nu}(w) \ran & = & -\frac{\alpha'}{2}\eta^{\mu\nu}(z-w)^{-1} \ ,\non \\
\lan\phi(z)\phi(w)\ran & = & -\log(z-w) \ .
\labell{prop}\end{eqnarray}

By applying  $x_{4}\equiv\ z=x+iy$ and $x_{5}\equiv\bz=x-iy$, one can get   the $C \phi TT$ amplitude as

\beqa {\cal A}^{C\phi TT}&\sim& 2\int
 dx_{1}dx_{2}dx_{3}dx_{4} dx_{5}\,
(P_{-}\fsH_{(n)}M_p)^{\al\be}\xi_{1i}k_{2a}(-\alpha'^2 k_{3b})x_{45}^{-1/4}(x_{14}x_{15})^{-1/2}I
\nonumber\\&&\times<:S_\al(x_4):S_\be(x_5):\psi^i(x_1):\psi^a(x_2):\psi^b(x_3)>\Tr(\lam_1\lam_2\lam_3),\labell{125}\eeqa where
\beqa
I&=&|x_{12}|^{\alpha'^2 k_1.k_2}|x_{13}|^{\alpha'^2 k_1.k_3}|x_{14}x_{15}|^{\frac{\alpha'^2}{2}k_1.p}|x_{23}|^{        \alpha'^2 k_2.k_3}|x_{24}x_{25}|^{  \frac{\alpha'^2}{2}  k_2.p}
|x_{34}x_{35}|^{ \frac{\alpha'^2}{2}   k_3.p}|x_{45}|^{ \frac{\alpha'^2}{4}  p.D.p}
\nonumber\eeqa

such that
$x_{ij}=x_i-x_j$ has been used. Applying the Wick-like rule~\cite{Liu:2001qa} and \cite{Garousi:2008ge,Hatefi:2010ik} to our case
 we end up with  the correlation function involving three fermions  ($\psi$s)
and two spin operators  as
\beqa
<:S_{\al}(x_4):S_{\be}(x_5):\psi^i(x_1):\psi^a(x_2)::\psi^b(x_3):>&=&
\bigg[(\Gamma^{bai}C^{-1})_{\alpha\beta}
-2\eta^{ab}\frac{Re[x_{24}x_{35}]}{x_{23}x_{45}}(\gamma^{i}C^{-1})_{\alpha\beta}\bigg]\nonumber\\&&\times 2^{-3/2}x_{45}^{1/4} (x_{14}x_{15}x_{24}x_{25}x_{34}x_{35})^{-1/2}
\label{esi1}\eeqa

Embedding \reef{esi1} into our amplitude, we were able to check the SL(2,R) invariance of the S-matrix of $C \phi TT$. We wish to fix the location of the open strings  in  special places, that is

 \beqar
 x_{1}=0 ,\qquad x_{2}=1,\qquad x_{3}\rightarrow \infty,
 \eeqar
 By using this particular gauge fixing, we get to the following integrals
\beqa
 \int d^2 \!z |1-z|^{a} |z|^{b} (z - \bar{z})^{c}
(z + \bar{z})^{d},
 \eeqa
where $d=0,1,2$ and $a,b,c$ should be given in terms of the Mandelstam variables as below

\beqar
s&=&-\frac{\alpha'}{2}(k_1+k_3)^2,\qquad t=-\frac{\alpha'}{2}(k_1+k_2)^2,\qquad u=-\frac{\alpha'}{2}(k_2+k_3)^2.
\qquad\eeqar

  The result of the integrations for just $d=0,1$ was expressed in  \cite{Fotopoulos:2001pt}, however if the computations were done in $C^{-2} \phi^{0} T^0 T^0$ picture we would need the result of the integrations for
 $d=2$ as well , which was performed in \cite{Hatefi:2012wj}.
\\

After  some computations now we write down the complete form of $C \phi TT$,
\beqa {\cal A}^{C\phi T T}&=&{\cal A}_{1}+{\cal A}_{2}\labell{181u}\eeqa

so that

\beqa
{\cal A}_{1}&\!\!\!\sim\!\!\!&-8\xi_{1i}k_{2a}k_{3b} 2^{-3/2}
\Tr(P_{-}\fsH_{(n)}M_p\Gamma^{bai}
)L_1,
\nonumber\\
{\cal A}_{2}&\sim&-8\xi_{1i} 2^{-3/2}\bigg\{\Tr(P_{-}\fsH_{(n)}M_p \gamma^{i})\bigg\}L_2
\labell{48}\eeqa

\vskip .1in

It should be emphasized that the closed form of $<V_C V_{\phi} V_T V_T>$ does not vanish only for $p=n$ and $p+2=n$ cases. As a matter of fact  we observe that ${\cal A}_{1}$ (first part of amplitude) in \reef{48} is antisymmetric under interchanging  two tachyons or under replacing $2\leftrightarrow 3$. This does show that  the first part of this four point function  for $p=n$ case should be vanished for both one RR and two real $T_1$ tachyons and one RR and two $T_2$'s .

\vskip .2in

 $L_1,L_2$ are made of
\beqa
L_1&=&(2)^{-2(t+s+u)-1}\pi{\frac{\Gamma(-u)
\Gamma(-s+\frac{1}{4})\Gamma(-t+\frac{1}{4})\Gamma(-t-s-u)}
{\Gamma(-u-t+\frac{1}{4})\Gamma(-t-s+\frac{1}{2})\Gamma(-s-u+\frac{1}{4})}},\nonumber\\
L_2&=&-(2)^{-2(t+s+u+1)}\pi{\frac{\Gamma(-u+\frac{1}{2})
\Gamma(-s+\frac{3}{4})\Gamma(-t+\frac{3}{4})\Gamma(-t-s-u-\frac{1}{2})}
{\Gamma(-u-t+\frac{1}{4})\Gamma(-t-s+\frac{1}{2})\Gamma(-s-u+\frac{1}{4})}}
\nonumber\eeqa

By talking about the correct momentum expansion we will actually understand  that how to
gain all infinite scalar, gauge poles for different $p,n$ values, how to interpret them in field theory and how to look for new couplings in the world volume of brane anti brane systems. We also explain why we do not have any tachyon pole in $<V_C V_{\phi} V_T V_T>$.

\vskip .1in

\section{Momentum expansion for brane-anti brane systems}

To be able to look for all couplings in string theory, we need to expand the complete form of the
S-matrix elements to find all singular terms (poles) and contact
interactions. However due to presence of tachyons here the expansion
is no longer the low energy expansion.  Therefore we are not allowed
to use  the limit of $\alpha'\rightarrow 0$ of the above string
amplitude. Turning into Mandelstam variables and using the momentum conservation along the world volume
of brane, $\alpha'(k_1^{a} +  k_2^{a}+ k_3^{a})+p^{a}+(p.D)^a =0$, we obtain a very useful constraint
as follows

\vskip .1in

 \beqa
s+t+u=-p_ap^a-\frac{1}{2}. \labell{cons}\eeqa

 In \cite{Hatefi:2012wj} we discussed in detail that the momentum expansion for various amplitudes should be found either by $(k_i+k_j)^2\rightarrow 0$
 or $k_i\inn k_j\rightarrow 0$.  The first case takes place when we do have a massless pole and for the other cases we should use $k_i\inn k_j\rightarrow 0$ expansion.
  Due to non-zero  two tachyon one gauge couplings, we realize that our amplitude $C\phi TT$ has to have  a massless  pole in $-(k_3+k_2)^2=u$ channel. It should only have infinite scalar poles in  $t'+s'+u$ channel $(t'=t+\frac{1}{4}, s'=s+\frac{1}{4})$ and it will become clear later on.

   \vskip .1in

   The extremely important point for brane -anti brane  systems is that the quantity $p^ap_a$ should tend to zero
while in non-BPS systems due to kinematic constraints  (which
Mandelstam variables justify) our amplitude like $C\phi\phi T$  \cite{Hatefi:2012wj} just makes sense for $p^ap_a \rightarrow \frac{1}{4}$.

   \vskip .1in

Keeping in mind above arguments, we understand that the correct momentum
expansion in brane anti brane system is  indeed

 \beqa
(k_3+k_2)^2\rightarrow 0,\qquad k_1.k_3\rightarrow 0,\qquad
k_2.k_1\rightarrow 0. \nonumber\eeqa

Also using the on-shell
relations $k_1^2=0$ and $k_2^2=k_3^2=\frac{1}{4}$
the above momentum expansion   can be interpreted in terms of Mandelstam variables as

  \beqa u\rightarrow 0,\qquad s\rightarrow
\frac{-1}{4},\qquad t\rightarrow \frac{-1}{4}\labell{point}\eeqa

 Therefore  $C \phi TT$ should be
performed just by setting  $p_ap^a\rightarrow 0$ in brane-anti
brane systems.

Applying
\reef{point} into the general form of the amplitude, we conclude the poles of the Gamma functions. Hence
our amplitude for  $p=n$ case, has infinite u-channel gauge poles
meanwhile it does have
 infinite scalar poles for $p+2=n$ case. Note that since the two tachyons, one scalar field coupling is vanished, we no longer have tachyon pole , rather than in $CATT$ amplitude for $p=n$ case we had infinite tachyon poles.


\vskip .1in

 Now we would like to expand our amplitude around \reef{point} to find new couplings out and also to gain
 the correct two tachyon two scalar couplings of brane anti brane systems to all orders in $\alpha'$.
 We also do want to confirm
all order corrections to the  Wess-Zumino effective actions of BPS branes which have recently been constructed in
\cite{Hatefi:2012ve,Hatefi:2012rx}. By deriving the closed form of the expansions, we search about  several new couplings  in field theory, set their coefficients precisely and eventually find all their infinite higher derivative corrections thereof.

\vskip .1in

\vskip .1in

\subsection{Infinite u-channel gauge poles  and  contact interactions for $p=n$ case}

The amplitude was anti symmetrized with respect to interchanging
two external tachyons therefore both four point amplitudes of  $C\phi T_1T_1$ and $C\phi T_2T_2$ make no sense in this case. By performing the trace, the ultimate form of  the amplitude of $C\phi T_1 T_2$ becomes
 \beqa
{\cal A}^{C\phi T_1T_2 }&=& -8
\xi_{1i}k_{3b}k_{2a}2^{-3/2}\frac{16}{(p)!}\left( \eps^{a_{0}\cdots
a_{p-2}ba}H^{i}_{a_{0}\cdots a_{p-2}} \right)L_1\labell{pnbb}\eeqa

\vskip .1in

Let us include the expansion of $L_1$ around \reef{point} at leading orders

 \beqa
L_1&=&
\pi^{3/2}\bigg(\frac{-1}{u}
+4\ln(2)+\bigg(\frac{\pi^2}{6}-8\ln(2)^2\bigg)u-\frac{\pi^2}{6}\frac{(s'+t')^2}{u}+\cdots\bigg)\labell{line} \eeqa

such that $s'=s+\frac{1}{4}, t'=t+\frac{1}{4}$.

\vskip .1in

It is clear from \reef{line} that the first term is pole, thus we have to produce this simple pole in field theory as well, although as we will see later this part of the S-matrix involves infinite gauge poles, however for the moment let us proceed order by order. In order to produce that gauge pole, we should have taken the following
 Feynman rule :

\vskip .1in

\beqa {\cal
A}&=&V_a(C_{p-1},\phi,A)G_{ab}(A)V_b(A,T_1,T_2)\labell{amp37}\eeqa
with

 \beqa G_{ab}(A) &=&\frac{ i\delta_{ab}}{(2\pi\alpha')^2 T_p
k^2}\nonumber\\
V_b(A,T_1,T_2)&=&T_p(2\pi\alpha')(k_{2b}-k_{3b})\label{aa1} \eeqa

Note that in the propagator $k^2$ is indeed $-\frac{\alpha'}{2}(k_2+k_3)^2=u$. To make field theory obvious, we just point out that the so called $\Tr(\frac{-1}{4} F_{ab}F^{ba})$(gauge field's kinetic term) is incharge of propagator.

 $V_b(A,T_1,T_2)$ has been found from  $\Tr(2\pi\alpha' D_aT D^aT)$ (tachyon's kinetic term), by applying direct field theory techniques.

 In order to obtain that pole we need to actually have the
coupling of  $V_a(C_{p-1},\phi,A)$ as follows:
 \beqa
 \mu_p\lambda (2\pi\alpha')\int_{\sum_{(p+1)}}\Tr\bigg(\partial_iC_{p-1}\wedge F (\phi_1+\phi_2)^i\bigg)\labell{newrr}
 \eeqa
 where $\lambda=2\pi\alpha'$ and we used the fact that amplitude is antisymmetric. The scalar field was found  from
Taylor expansion. Note that this coupling without $\phi_2^i$ has been firstly pointed out in \cite{Hatefi:2012ve} and it was obtained by carrying out direct computation of $C \phi AA$ amplitude in the world volume of BPS branes.
Having set \reef{newrr}, we get to

\beqa
V_a(C_{p-1},\phi,A
)&=&\mu_p(2\pi\alpha')^2\frac{1}{(p)!}\epsilon_{a_0\cdots a_{p-2}
}H^{ia_0\cdots
a_{p-2}} k_{a}\xi_{i}\nonumber\eeqa

with $k$ becomes  the off-shell gauge field's momentum and it is $k_a=-(k_2+k_3)_a$.
It is not difficult to indeed show that by normalizing the amplitude with a coefficient of
$\frac{i\mu_p}{4\sqrt{2\pi}}$ and using \reef{amp37} the first gauge pole in \reef{line} will precisely be resulted.

\vskip .1in

Now we come to interesting point which is discovering new couplings. As it is clear from the expansion, the second term in \reef{line} is contact term so in order to produce this term the following coupling should have been  considered

\beqa
\frac{i}{2} \mu_p \beta^2 (2\pi\alpha')^3  \Tr\bigg(\partial_{i} C_{p-1}\wedge DT \wedge DT^{*}(\phi^1+\phi^2)^i\bigg)
\label{nnew}
\eeqa
Thus by extracting this new Wess-Zumino coupling as follows

\beqa
A_c&=&\mu_p \beta^2 (2\pi\alpha')^3 \frac{i}{2p!}\epsilon_{a_0\cdots a_{p}}H^{ia_0\cdots a_{p-2}} k^{3a_{p-1}}  k^{2a_{p}}\xi^{i}
\label{ct1}
\eeqa
we could exactly produce the second term of the expansion. Notice that $\beta$ is a normalization constant in Wess-Zumino  terms which has been fixed in \cite{Garousi:2007fk}.

\vskip .1in

Again the third term of the expansion is contact interaction. A very important question which
comes out is that how we can  talk about the other terms appeared in \reef{line}. The simple expectation is that the other terms must be related to higher derivative corrections of the new couplings so having applied this remarkable argument, we are led to the following coupling

\beqa
\frac{i}{2} \mu_p  (2\pi\alpha') (\alpha')^2 \bigg(\frac{\pi^2}{6}-8 ln2^2\bigg) \Tr\bigg(\partial_{i} C_{p-1}\wedge D^aD_a( DT \wedge DT^{*})(\phi^1+\phi^2)^i\bigg)
\label{nnew22}
\eeqa

Having used the higher derivative correction to the coupling of

\beqa \Tr\bigg(\partial_{i} C_{p-1}\wedge DT \wedge DT^{*}(\phi^1+\phi^2)^i\bigg)\nonumber\eeqa

 as appeared in \reef{nnew22}, we could produce the third term of the expansion in \reef{line} in a correct manner, so not only could we find new coupling like \reef{nnew} but also we could fix its coefficient, more interestingly  we could obtain its higher derivative extensions in \reef{nnew22} without any on-shell ambiguity.

\vskip .2in

The other fascinating fact which we must emphasize is that, the expansion \reef{line} dictates us that our S-matrix has namely infinitely many gauge poles. For example the fourth term in the expansion of \reef{line} is again simple pole. All the infinite gauge poles are
going to be produced by the same Feynman rule \reef{amp37} but two subtleties
are in order. First of  all the whole infinite gauge poles
will show us that the term of  $\Tr(2\pi\alpha' D_aT D^aT$) or in the other
words,  tachyon's kinetic term does not  receive any higher derivative
correction. This is related to the fact that this kinetic term or $\Tr(2\pi\alpha' D_aT D^aT$) has been fixed
in tachyon effective action. Second of all, the natural point comes out is that, the only approach to obtain all infinite gauge poles is to devote the  higher derivative corrections to the vertex of $V_a(C_{p-1},A,\phi^i)$. Thus it must get improved  to be able to produce all infinite poles  in field theory analysis as well.

\vskip .1in

For example the second pole for this amplitude is produced by embedding the following vertex in the Feynman amplitude \reef{amp37}

\beqa
V_a(C_{p-1},\phi,A
)&=& \frac{(\alpha' \pi)^2}{6}\mu_p(2\pi\alpha')^2\frac{1}{(p)!}\epsilon_{a_0\cdots a_{p-2}
}H^{ia_0\cdots
a_{p-2}} k_{a}\xi_{i} (k_1.p)^2\nonumber\eeqa
Now in order to produce all infinite gauge poles, the complete form of the expansion of $L_1$ is needed. After applying some identities, the closed form of the expansion of
 $L_1$ around \reef{point} to all orders in $\alpha'$ can be explored as follows

\beqa
L_1&=&
\pi^{3/2}\left(-\frac{1}{u}\sum_{n=-1}^{\infty}b_n(s'+t')^{n+1}+\sum_{p,n,m=0}^{\infty}f_{p,n,m}u^p\left(s't'\right)^n(s'+t')^m\right)
\label{ee1}\eeqa

with $f_{p,0,0}=a_p$ and the following coefficients must be considered

  \beqa
&&b_{-1}=1,\,b_0=0,\,b_1=\frac{\pi^2}{6},\,b_2=2\z(3),a_0=4ln2,\nonumber\\&&
a_1=\frac{\pi^2}{6}-8ln(2)^2,a_2=\frac{2}{3}(-\pi^2 ln2+3\z(3)+16ln(2)^3),\nonumber\\
&&f_{0,0,2}=\frac{2}{3}\pi^2\ln(2),\,f_{0,1,0}=-14\z(3),
f_{0,0,3}=8\z(3)\ln(2),\nonumber\\
&&f_{1,1,0}=56\z(3)\ln(2)-1/2,\,f_{1,0,2}=\frac{1}{36}(\pi^4-48\pi^2\ln(2)^2),\,f_{0,1,1}=-1/2\label{rre}
\eeqa

Note that some of these coefficients are completely different from those appeared in the world volume of non-BPS branes  \cite{Hatefi:2012wj}. However it is interesting that $b_n$ coefficients of this paper  are precisely the same $b_n$'s that were appeared in the momentum expansion of BPS branes ( just like $b_n$ coefficients in the S-matrix elements of $CAAA$ in \cite{Hatefi:2010ik}).

By imposing the infinite higher derivative corrections as

 \beqa
 \mu_p\lambda (2\pi\alpha')\int_{\sum_{(p+1)}}\partial_iC_{p-1}\wedge  \sum_{n=-1}^{\infty}b_{n}(\alpha')^{n+1} \Tr\bigg(
   D_{a_{1}}\cdots D_{a_{n+1}}F D^{a_{1}}\cdots D^{a_{n+1}} (\phi_1+\phi_2)^i\bigg)\labell{new287}
 \eeqa

 deriving the vertex of
$V_a(C_{p-1},\phi,A)$ as

 \beqa
V_a(C_{p-1},\phi,A)&=&\mu_p(2\pi\alpha')^2\frac{1}{(p)!}\epsilon_{a_0\cdots a_{p-2}
}H^{ia_0\cdots
a_{p-2}}(k_2+k_3)^{a}\xi_{i}\sum_{n=-1}^{\infty}b_n(\alpha'k_1\cdot
k)^{n+1}\nonumber\eeqa

and replacing it in \reef{amp37} we end up with

\beqa {\cal
A}=\mu_p(2\pi\alpha')\frac{2i}{(p)!u}\epsilon_{a_0\cdots
a_{p-2}ab}H^{ia_0\cdots
a_{p-2}}k_2^{b}k_3^{a}\xi_i\sum_{n=-1}^{\infty}b_n\left(\frac{\alpha'}{2}\right)^{n+1}(s'+t')^{n+1}\labell{AAbb}\eeqa

By replacing the first term of \reef{ee1} in \reef{pnbb} and comparing \reef{AAbb} with \reef{pnbb} we actually clarified that  the infinite  string theory 'gauge poles are  precisely produced. The other remarkable fact which should be emphasized is that,by comparing the above amplitude \reef{pnbb} with the infinite gauge
poles of the field theory amplitude, that is, \reef{AAbb}  we have no longer any residual contact interactions for  $p=n$ case.

\vskip .2in

Let us end this part of the amplitude by searching about all contact terms of $C \phi TT$ amplitude as below

\beqa
-\frac{i}{2}\mu_p \pi \xi_{1i}k_{3b}k_{2a} \frac{16}{(p)!}\left( \eps^{a_{0}\cdots
a_{p-2}ba}H^{i}_{a_{0}\cdots a_{p-2}} \right)\sum_{p,n,m=0}^{\infty}f_{p,n,m}u^p\left(s't'\right)^n(s'+t')^m \labell{pncc1}\eeqa

Working out in detail we can show that  the closed form of all infinite  contact terms of string theory
amplitude for $p=n$ case,  will be concluded in the field theory  by taking into account the following couplings to all orders in $\alpha'$ :

\beqa
&&2i\alpha'(\pi\alpha')\mu_p\sum_{p,n,m=0}^{\infty}f_{p,n,m}\left(\frac{\alpha'}{2}\right)^{p}
\left(\alpha'\right)^{2n+m} \partial_i C_{p-1} \wedge(D^aD_a)^p  D_{b_1}\cdots
D_{b_{m}}(D_{a_1}\cdots D_{a_n}DT\nonumber\\
&&\wedge D_{a_{n+1}}\cdots
D_{a_{2n}}DT^*)\prt^{a_1}\cdots\prt^{a_{2n}}\prt^{b_1}\cdots\prt^{b_{m}}(\phi_1+\phi_2)^i\labell{hderv1n2}
\eeqa

Note that in the above coupling we could consider covariant derivative of scalar, however it turns out that the commutator in the definition of covariant derivative of scalar should not be considered here as we are looking for $C \phi TT$ coupling. A  good question is to see whether or not covariant derivative of scalar keeps fixed in the above coupling. To answer this question one has to carry out a six point function, basically checking $C \phi A TT$ amplitude will be worth, nevertheless we can not come over to this remark by our computations of  this paper.

\subsection{Infinite $(u+s'+t')$-channel scalar poles
 and  contact
  interactions for $p+2=n$ case}

Having performed the trace, we can take all non-zero terms of the string amplitude for this certain case as

\beqa
{\cal A}^{\phi T_1T_1C}&\!\!\!\!=\!\!\!\!&\pm\frac{8i \mu_p}{\sqrt{\pi}(p+1)!}L_2  \eps^{a_{0}\cdots a_{p}}H^i_{a_{0}\cdots a_{p}}   \xi_i \labell{wwmm}\eeqa

 As it follows from the amplitude and unlike the first part of the amplitude, in this section our S-matrix is symmetric, that is, by  interchanging two tachyons the amplitude remains unchanged. Therefore $C \phi T_1 T_2$ amplitude does not have any contribution to $p+2=n$ case. As a matter of fact our calculations make sense either for $C \phi T_1 T_1$ or $C \phi T_2 T_2$ amplitudes. Note that due to kinematic reasons, we can not talk about associated ward identities either.

\vskip .1in

The expansion of $L_2$ at leading order  around \reef{point} is

\beqa
L_2&=&\frac{\sqrt{\pi}}{2}\bigg(\frac{-1}{(t'+s'+u)}+ 4\ln(2)+\bigg(\frac{\pi^2}{6}-8\ln(2)^2\bigg)(s'+t'+u)\nonumber\\&&-\frac{\pi^2}{3}\frac{t's'}{(t'+s'+u)}+\cdots\bigg)\labell{high22}\eeqa

 The first scalar pole in $t'+s'+u$-channel should be produced by taking the following amplitude

\beqa
{\cal A}&=&V_i(C_{p+1},\phi^{(1)})G_{ij}(\phi)V_j(\phi^{(1)},T_1,T_1,\phi^{(1)})\nonumber\\&&
+V_i(C_{p+1},\phi^{(2)})G_{ij}(\phi)V_j(\phi^{(2)},T_1,T_1,\phi^{(1)})\labell{amp44}\eeqa

It is important to highlight the fact that in order to have consistent result  $\phi$ in the propagator must be $\phi^{(1)}$ and $\phi^{(2)}$. The needed vertices can be read off as

\beqa
G_{ij}(\phi) &=&\frac{i\delta_{ij} \delta_{\alpha\beta}}{(2\pi\alpha')^2 T_p
\left(u+t'+s'\right)}\nonumber\\
V_i(C_{p+1},\phi^{(1)})&=&i\mu_p(2\pi\alpha')\frac{1}{(p+1)!}\epsilon^{a_0\cdots a_{p}}H_{ia_0\cdots a_{p}}\Tr(\lambda_\alpha)\nonumber\\
V_i(C_{p+1},\phi^{(2)})&=&-i\mu_p(2\pi\alpha')\frac{1}{(p+1)!}\epsilon^{a_0\cdots a_{p}}H_{ia_0\cdots a_{p}}\Tr(\lambda_\alpha)\nonumber\\
V_j(\phi^{(1)},T_1,T_1,\phi^{(1)})&=&-2iT_p(2\pi\alpha')\xi_j \Tr(\lambda_1\lambda_2\lambda_3\lambda_\beta)\nonumber\\
V_j(\phi^{(2)},T_1,T_1,\phi^{(1)})&=&2iT_p(2\pi\alpha')\Tr(\lambda_1\lambda_2\lambda_3\lambda_\beta)\xi_j\labell{Fey}
\eeqa

Now by applying the above vertices into  \reef{amp44}  we reach to

\beqa
{\cal A}&=&\frac{4i\mu_p}{(p+1)!(u+s'+t')}\eps^{a_{0}\cdots a_{p}}H_{ia_{0}\cdots a_{p}}\xi^i\labell{amp6}\eeqa

By substituting \reef{high22} into \reef{amp44} and comparing  the result with \reef{amp6} we get the fact that  the first scalar pole in $u+t'+s'$-channel has been precisely obtained.
\vskip .2in

From this part of the amplitude we understand that the presence of

\beqa
D\phi^{(1)i}.D\phi^{(2)}_i\label{mmbk}\eeqa

 term in the effective action is inevitable. In fact if we do not consider $D\phi^{(1)i}.D\phi^{(2)}_i$ term in the effective action
 then
 $V_j(\phi^{(2)},T_1,T_1,\phi^{(1)})$ does not have any value and naturally  we are not able  to produce even the first simple scalar pole in field theory where this new term \reef{mmbk} is not inside of ordinary trace effective action.
\vskip .2in

Let us consider the first contact interaction in \reef{high22} as follows
\beqa
{\cal A}_c&=&i\mu_p\ln(2)\frac{16}{(p+1)!} \eps^{a_{0}\cdots a_{p}}H_{ia_{0}\cdots a_{p}}\xi^i\labell{amp61}\eeqa

It is remarkable to point out the following point. In order to produce \reef{amp61}, we should explore another new coupling and indeed fix its coefficient, which is the only way to do so as

\beqa
\frac{\mu_p}{(p+1)!}(2\pi\alpha')^2\beta^2  \partial_i C_{a_{0}\cdots a_{p}}(\phi^{(1)}-\phi^{(2)})^i|T|^2 \eps^{a_{0}\cdots a_{p}}
\label{nc2}
\eeqa

  We believe that the third term in the expansion of \reef{high22},  will be produced just by applying higher derivative on the above coupling as
\beqa
-(\alpha')^2\frac{\mu_p}{(p+1)!}\left(\frac{\pi^2}{6}-8\ln(2)^2\right) \partial_i C_{a_{0}\cdots a_{p}} D^aD_a\bigg[(\phi^{(1)}-\phi^{(2)})^i|T|^2\bigg]\eps^{a_{0}\cdots a_{p}}\labell{hderv25}
\eeqa

Another way of writing this higher extension is as follows :

\beqa
-(\alpha')^2\frac{\mu_p}{(p+1)!}\left(\frac{\pi^2}{6}-8\ln(2)^2\right)
\eps^{a_{0}\cdots a_{p}}H_{ia_{0}\cdots a_{p}}
  \partial^a\partial_a \bigg[(\phi^{(1)}-\phi^{(2)})^i TT^*\bigg]\label{ra1}\eeqa

 One important point has to be  made. This coupling and in particular the difference between the first scalar (which is on the brane) and the second scalar (which is on the anti brane) inside this new coupling \reef{ra1} has to appear, to be able to produce the third term of the expansion in \reef{high22}, otherwise we get inconsistent result.

\vskip .2in

Likewise the last section , we are going to find out the closed form of the expansion of $L_2$  around \reef{point} as
\beqa
L_2&=&\frac{\sqrt{\pi}}{2}\bigg(\frac{-1}{(t'+s'+u)}+ \sum_{n=0}^{\infty}a_n(s'+t'+u)^n+\frac{\sum_{n,m=0}^{\infty}l_{n,m}(s'+t')^n(t's')^{m+1}}{(t'+s'+u)}\nonumber\\&&+\sum_{p,n,m=0}^{\infty}e_{p,n,m}(s'+t'+u)^p(s'+t')^n(t's')^{m+1}\bigg)\label{mmn}\eeqa

$l_{n,m}$ and $e_{p,n,m}$ are
 \beqa
l_{0,0}=-\pi^2/3,\,&&l_{1,0}=8\z(3)\\
l_{2,0}=-7\pi^4/45,\, l_{0,1}=\pi^4/45,\,&&\,l_{3,0}=32\z(5),\, l_{1,1}=-32\z(5)+8\z(3)\pi^2/3\nonumber\\
e_{0,0,0}=\frac{2}{3}\left(2\pi^2\ln(2)-21\z(3)\right),&& e_{1,0,0}=\frac{1}{9}\left(4\pi^4-504\z(3)\ln(2)+24\pi^2\ln(2)^2\right)\nonumber
\eeqa

\vskip .1in

It is highly important to trust this special expansion in order to distinguish  the last contact terms of \reef{mmn} from the second terms of expansion in \reef{mmn}. Indeed the last terms in \reef{mmn} do have different structure so we come to the fact that the last term of the expansion must be produced by making use of the different couplings in field theory.


 \vskip .2in

 Using the arguments  we have mentioned earlier, we can write down a higher derivative coupling to produce the second term in \reef{mmn}. In the other words  it is produced by taking this higher extension of the coupling

 \beqa
-(\alpha')^2\frac{\mu_p}{(p+1)!}\sum_{n=0}^{\infty}a_n  \left(\frac{\alpha'}{2}\right)^n
\eps^{a_{0}\cdots a_{p}}H_{ia_{0}\cdots a_{p}}
   (\partial^a\partial_a)^n \bigg[(\phi^{(1)}-\phi^{(2)})^i TT^*\bigg]\eeqa

  Remember that we found these couplings with using the S-matrix of  string theory,
   thus we believe that here all  on-shell ambiguities for these new couplings are removed.

\vskip .2in

As it is clear from the closed form of the expansion of $L_2$ in \reef{mmn}, it does involve the infinite
 scalar poles in $t'+s'+u$-channel for $p+2=n$ case. Let us write down in a precise way all the infinite scalar poles in string theory once more:

   \beqa {\cal A}^{C\phi
T_2T_2}&\!\!\!\!=\!\!\!\!& \frac{4i\mu_p}{ (p+1)!}
\eps^{a_{0}\cdots a_{p}}H_{ia_{0}\cdots a_{p}}
\bigg(\xi^i s't' \bigg)\frac{\sum_{n,m=0}^{\infty}l_{n,m}(s'+t')^n(t's')^{m}}{(t'+s'+u)}\labell{pniy}\eeqa

In this subsection, we want to obviously show that the recent all order two scalar two tachyon
couplings of non-BPS branes (equation (43) of \cite{Hatefi:2012wj}) will not produce
infinite massless scalar poles  of brane anti brane 's string theory amplitude of $C \phi T_2 T_2$ which exist
  in   $(s'+t'+u)$-channel.
\vskip .1in

   Thus we conclude that, one has to discover  the new all order higher derivative corrections of two tachyon, two scalar couplings of  brane anti brane systems. Finally in order to show that we have found the correct corrections to two tachyon two scalar couplings to all orders in brane anti brane systems  we make a consistent check by producing all infinite scalar poles of our S-Matrix ($C\phi TT$).

    In fact by making use of these new higher derivative corrections we exactly produce all the infinite  scalar poles in $u+t'+s'$ channel. Although we do not explain all details for corrections ,  we just make a comment  that the appearance of $\sigma_3$ Chan-Paton factor for scalar field in (-1)-picture plays a crucial role  to actually obtain these corrections (more information can be found in \cite{Hatefi:2012wj,Hatefi:2012rx}).

$\frac{}{}$

 The  Lagrangian for two scalar, two tachyon couplings for non-BPS brane was defined as
\beqa
{\cal{L}}(\phi,\phi,T,T)&=& -2T_p(\pi\alpha')^3{\rm
STr} \bigg(
m^2\cT^2(D_a\phi^iD^a\phi_i)+\frac{\alpha'}{2}D^{\alpha}\cT D_{\alpha}\cT D_a\phi^iD^a\phi_i \nonumber\\&&-
\alpha' D^{b}\cT D^{a}\cT D_a\phi^iD_b\phi_i \bigg)\labell{dbicoupling} \eeqa

 Having extracted symmetrized traces, the higher derivative corrections of two scalars, two tachyons for non-BPS branes   to all orders in $\alpha'$  have been found in \cite{Hatefi:2012wj} as

\beqa
\cL_{}&=&-2T_p(\pi\alpha')(\alpha')^{2+n+m}\sum_{n,m=0}^{\infty}(\cL_{1}^{nm}+\cL_{2}^{nm}+\cL_{3}^{nm}+\cL^{nm}_{4}),\labell{lagrango}\eeqa
where
\beqa
\cL_1^{nm}&=&m^2
\Tr\left(\frac{}{}a_{n,m}[\cD_{nm}(\cT^2 D_a\phi^iD^a\phi_i)+ \cD_{nm}(D_a\phi^iD^a\phi_i\cT^2)]\right.\nonumber\\
&&\left.+\frac{}{}b_{n,m}[\cD'_{nm}(\cT D_a\phi^i\cT D^a\phi_i)+\cD'_{nm}( D_a\phi^i\cT D^a\phi_i\cT)]+h.c.\right),\nonumber\\
\cL_2^{nm}&=&\Tr\left(\frac{}{}a_{n,m}[\cD_{nm}(D^{\alpha}\cT D_{\alpha}\cT D_a\phi^iD^a\phi_i)+\cD_{nm}( D_a\phi^iD^a\phi_i D^{\alpha}\cT D_{\alpha}\cT)]\right.\nonumber\\
&&\left.+\frac{}{}b_{n,m}[\cD'_{nm}(D^{\alpha} \cT D_a\phi^i D_{\alpha}\cT D^a\phi_i)+\cD'_{nm}( D_a\phi^i D_{\alpha}\cT D^a\phi_i D^{\alpha} \cT)]+h.c.\right),\nonumber\\
\cL_3^{nm}&=&-\Tr
\left(\frac{}{}a_{n,m}[\cD_{nm}(D^{\beta}\cT D_{\mu}\cT D^\mu\phi^iD_\beta\phi_i)+\cD_{nm}( D^\mu\phi^iD_\beta\phi_iD^{\beta}\cT D_{\mu}\cT)]\right.\nonumber\\
&&\left.+\frac{}{}b_{n,m}[\cD'_{nm}(D^{\beta}\cT D^\mu\phi^iD_{\mu}\cT D_\beta\phi_i)+\cD'_{nm}(D^\mu\phi^i D_{\mu}\cT  D_\beta\phi_i  D^{\beta}\cT)]+h.c.\right),\nonumber\\
\cL_4^{nm}&=&-\Tr\left(\frac{}{}a_{n,m}[\cD_{nm}(D^{\beta}\cT D^{\mu}\cT D_\beta\phi^iD_\mu\phi_i)
+\cD_{nm}( D^\beta\phi^iD^\mu\phi_iD_{\beta}\cT D_{\mu}\cT)]\right.\nonumber\\
&&\left.+\frac{}{}b_{n,m}[\cD'_{nm}(D^{\beta}\cT D_\beta\phi^iD^{\mu}\cT D_\mu\phi_i)+\cD'_{nm}( D_\beta\phi^i D_{\mu}\cT  D^\mu\phi_i D^{\beta}\cT)]+h.c.
\right)\label{hdts}
\eeqa

Note that the definitions of $\cD_{nm}(EFGH)$ and $\cD'_{nm}(EFGH)$ were given in \cite{Hatefi:2012wj}.
 Let us show in a clear way that these corrections do not work for brane anti brane systems, namely we want to emphasize that by using them we can not produce all the infinite scalar poles in $u+t'+s'$-channel. Thus it definitely means that they are not correct corrections for brane anti brane systems and one has to discover new corrections for two tachyons and two scalars of brane anti brane system and distinguish them from the corrections of two tachyons and two scalars of non-BPS branes, which is one of the main goals of this paper.
\vskip .1in

 The Feynman rule for this particular case is

\beqa {\cal
A}&=&V_{\alpha}^{i}(C_{p+1},\phi)G_{\alpha\beta}^{ij}(\phi)V_{\beta}^{j}(\phi,\phi_1,
T_2,T_2),\labell{amp5419} \eeqa
with
\beqa G_{\alpha\beta}^{ij}(\phi)
&=&\frac{-i\delta_{\alpha\beta}\delta^{ij}}{T_p(2\pi\alpha')^2
k^2}=\frac{-i\delta_{\alpha\beta}\delta^{ij}}{T_p(2\pi\alpha')^2
(t'+s'+u)},\nonumber\\
V_{\alpha}^{i}(C_{p+1},\phi)&=&i(2\pi\alpha')\mu_p\frac{1}{(p+1)!} \eps^{a_0\cdots
a_{p}}
 H^{i }_{a_0\cdots a_{p}}\Tr(\lambda_{\alpha}).
\labell{Fey4} \eeqa

  $\lambda_{\alpha}$ is an Abelian matrix. Having considered  off-shell 's scalar field which is Abelian and taking two permutations as
\beqa \Tr(\lam_2\lam_3\lam_1\lambda_{\beta}),
\Tr(\lam_2\lam_3\lambda_{\beta}\lam_1 ) \nonumber\eeqa

  $ V_{\beta}^{j}(\phi,\phi_1,T_2,T_2)$  should be derived from the higher derivative couplings in \reef{hdts} as

\beqa V_{\beta}^{j}(\phi,\phi_1, T_2,T_2)&=&\xi_{1}^j I_9 (-2i
T_p\pi)(\alpha')^{n+m+3}(a_{n,m}+b_{n,m}) \bigg(\frac{}{}(k_2\inn
k_1)^n(k\inn k_2)^m+(k_2\inn k_1)^n(k_3\inn k_1)^m
\nonumber\\&&+(k_2\inn k)^n(k_1\inn k_2)^m+(k\inn k_2)^n (k\inn
k_3)^m  +(k_3\inn k)^n(k_2\inn k)^m+(k_3\inn k)^n(k_3\inn
k_1)^m\nonumber\\&& +(k_1\inn k_3)^n(k_2\inn k_1)^m+(k_3\inn
k_1)^n(k_3\inn k)^m \bigg),\labell{verpptt}\eeqa

 like previous notations  $k$  becomes
 off-shell scalar field's momentum, and
\beqa I_9&=&
\Tr(\lam_1\lam_2\lam_3\lambda_{\beta})\frac{1}{2}(s')(t')\eeqa

 $b_{n,m}$ is
symmetric, see \cite{Hatefi:2010ik}.
Although the coefficients have been mentioned in \cite{Hatefi:2012wj}, listing some of the coefficients like
$a_{n,m}$ and $b_{n,m}$ is important to address :
\beqa
&&a_{0,0}=-\frac{\pi^2}{6},\,b_{0,0}=-\frac{\pi^2}{12},a_{1,0}=2\z(3),\,a_{0,1}=0,\,b_{0,1}=-\z(3),a_{1,1}=a_{0,2}=-7\pi^4/90,\nonumber\\
&&a_{2,2}=(-83\pi^6-7560\z(3)^2)/945,b_{2,2}=-(23\pi^6-15120\z(3)^2)/1890,a_{1,3}=-62\pi^6/945,\nonumber\\
&&\,a_{2,0}=-4\pi^4/90,\,b_{1,1}=-\pi^4/180,\,b_{0,2}=-\pi^4/45,a_{0,4}=-31\pi^6/945,a_{4,0}=-16\pi^6/945,\nonumber\\
&&a_{1,2}=a_{2,1}=8\z(5)+4\pi^2\z(3)/3,\,a_{0,3}=0,\,a_{3,0}=8\z(5),b_{1,3}=-(12\pi^6-7560\z(3)^2)/1890,\nonumber\\
&&a_{3,1}=(-52\pi^6-7560\z(3)^2)/945, b_{0,3}=-4\z(5),\,b_{1,2}=-8\z(5)+2\pi^2\z(3)/3,\nonumber\\
&&b_{0,4}=-16\pi^6/1890.\eeqa
One needs to use the following relations as well

\beqa
  k_3\inn k&=&k_2.k_1-k^2, \quad\quad k_2\inn k=k_1.k_3-k^2 \nonumber\eeqa

 By ignoring some of the contact interactions (we go through them in the next section), we get all poles in field theory as
\beqa
&&8i\mu_p\frac{\eps^{a_{0}\cdots a_{p}}\xi_{i} H^{i}_{a_0\cdots
a_{p}}}{(p+1)!(s'+t'+u)}\Tr(\lam_1\lam_2\lam_3)
\sum_{n,m=0}^{\infty}(a_{n,m}+b_{n,m})[s'^{m}t'^{n}+s'^{n}t'^{m}]
  s't' \label{amphigh8}\eeqa

On the other hand all the infinite scalar poles of the  amplitude in string theory are given  as

\beqa
{\cal A}^{\phi_1T_2T_2C}&\!\!\!\!=\!\!\!\!&\pm\frac{4i \mu_p}{(p+1 )!}\eps^{a_{0}\cdots a_{p}}H^i_{a_{0}\cdots a_{p}} \frac{\sum_{n,m=0}^{\infty}l_{n,m}(s'+t')^n(t's')^{m+1}}{(t'+s'+u)}  \xi_i
\labell{ww}\eeqa

Now we explicitly show that the corrections of non-BPS branes  in \reef{hdts} can not be used for brane anti brane systems. After removing the common coefficients from both string and field theory amplitude, we are going to compare both string and field theory amplitude at zero and first order of $\alpha'$ such that
at zero order in field theory we get
\beqa
16s't'(a_{0,0}+b_{0,0})&=&-4\pi^2 s't'\nonumber\eeqa

while in string theory we get

\beqa
4l_{0,0} s't'&=&\frac{-4\pi^2}{3} s't'=4 l_{0,0} s't'\label{bvc}\eeqa
 Since these two coefficients are not the same , clearly it shows that we can not make those corrections of non-BPS  branes \reef{hdts} to brane anti brane systems.
 Let us also see what happens at first order of $\alpha'$ in both string and field theory sides.

\vskip .1in
At $\alpha'$ order, the field theory amplitude has

\beqa
8 s't'(s'+t')(a_{1,0}+a_{0,1}+b_{0,1}+b_{0,1})&=&0\nonumber\eeqa

 while in string theory at first order we get
 \beqa
 4l_{1,0} s't'(s'+t')&=&32 \z(3)s't'(s'+t')\nonumber\eeqa

Thus even at first order of $\alpha'$ it certainly confirms that \reef{hdts}  gives us the wrong result for brane anti brane systems. Therefore we come to important fact that one has to find out the correct higher derivative corrections of brane anti brane systems to all orders of $\alpha'$ in order to be able to produce all infinite massless scalar poles in $t'+s'+u$- channel. In the next section we construct them and check them out for all orders of $\alpha'$.

\subsection{  Higher derivative corrections to two tachyon-two scalar couplings for brane anti brane systems to all orders in $\alpha'$ }

In this section we are going to propose  higher derivative corrections to two tachyons, two scalar field couplings in brane anti brane systems to all orders in  $\alpha'$. As it is obvious the action for brane anti brane system might be constructed  by making use of the projection on the effective actions of two unstable branes. Indeed the projection is $(-1)^{F_L}$ where $F_L$ becomes space time 's left handed fermion number. Details for constructing higher derivative corrections to all orders in $\alpha'$, for  BPS branes
in \cite{Hatefi:2010ik,Hatefi:2012ve}, for non-BPS branes in \cite{Hatefi:2012wj,Garousi:2008ge,Hatefi:2012rx} and for brane anti brane systems namely for the couplings between two  tachyons and two gauge fields  in \cite{Garousi:2007fk} were given.

 \vskip .2in

  In order to avoid some details we just make a very important point as follows. To derive all the higher derivative corrections it turns out that the internal degree of freedom of the scalar field in (-1)-picture
   (which is $\sigma_3$) plays the crucial role.


By substituting the correct Chan-Paton matrices and extracting the related traces (for more information see \cite{Hatefi:2012rx}), we end up indeed with two scalar field and two tachyon couplings to all orders of $\alpha'$ for brane anti brane system as below:

\beqa
\cL_{}&=&-2T_p(\pi\alpha')(\alpha')^{2+n+m}\sum_{n,m=0}^{\infty}(\cL_{1}^{nm}+\cL_{2}^{nm}+\cL_{3}^{nm}+\cL^{nm}_{4}),\labell{lagrango212}\eeqa
where
\beqa
\cL_1^{nm}&=&m^2
\Tr\left(\frac{}{}a_{n,m}[\cD_{nm}(T T^* D_a\phi^{(1)i}D^a\phi^{(1)}_i)+ \cD_{nm}( D_a\phi^{(1)i}D^a\phi^{(1)}_i  T T^*)+h.c.]\right.\nonumber\\
&&\left.-\frac{}{}b_{n,m}[\cD'_{nm}(\cT D_a\phi^{(2)i}\cT^* D^a\phi^{(1)}_i)+\cD'_{nm}( D_a\phi^{(1)i}\cT D^a\phi^{(2)}_i \cT^*)+h.c.]\right),\nonumber\\
\cL_2^{nm}&=&\Tr\left(\frac{}{}a_{n,m}[\cD_{nm}(D^{\alpha}\cT D_{\alpha}\cT^* D_a\phi^{(1)i}D^a\phi^{(1)}_i)+\cD_{nm}( D_a\phi^{(1)i}D^a\phi^{(1)}_i D^{\alpha}\cT D_{\alpha}\cT^*)+h.c.]\right.\nonumber\\
&&\left.-\frac{}{}b_{n,m}[\cD'_{nm}(D^{\alpha} \cT D_a\phi^{(2)i} D_{\alpha}\cT^* D^a\phi^{(1)}_i)+\cD'_{nm}( D_a\phi^{(1)i} D_{\alpha}\cT D_a\phi^{(2)}_i D^{\alpha} \cT^*)+h.c.]\right),\nonumber\\
\cL_3^{nm}&=&-\Tr
\left(\frac{}{}a_{n,m}[\cD_{nm}(D^{\beta}\cT D_{\mu}\cT^* D^\mu\phi^{(1)i} D_\beta\phi^{(1)}_i)+\cD_{nm}( D^\mu\phi^{(1)i} D_\beta \phi^{(1)}_iD^{\beta}\cT D_{\mu}\cT^*)+h.c.]\right.\nonumber\\
&&\left.-\frac{}{}b_{n,m}[\cD'_{nm}(D^{\beta}\cT D^\mu\phi^{(2)i}D_{\mu}\cT^* D_\beta\phi^{(1)}_i)+\cD'_{nm}(D^\mu\phi^{(1)i} D_{\mu}\cT  D_\beta\phi^{(2)}_i D^{\beta}\cT^*)+h.c.]\right),\nonumber\\
\cL_4^{nm}&=&-\Tr\left(\frac{}{}a_{n,m}[\cD_{nm}(D^{\beta}\cT D^{\mu}\cT^* D_\beta\phi^{(1)i}D_\mu\phi^{(1)}_i)
+\cD_{nm}( D^\beta\phi^{(1)i}D^\mu\phi^{(1)}_iD_{\beta}\cT D_{\mu}\cT^*)+h.c.]\right.\nonumber\\
&&\left.-\frac{}{}b_{n,m}[\cD'_{nm}(D^{\beta}\cT D_\beta\phi^{(2)i}D^{\mu}\cT^* D_\mu\phi^{(1)}_i)+\cD'_{nm}( D_\beta\phi^{(1)i} D_{\mu}\cT  D^\mu\phi^{(2)}_iD^{\beta}\cT^*)+h.c.]
\right)\label{hdts212}
\eeqa

In addition to the above couplings , one has to interchange  $ D\phi^{(1)i} $ to $ D\phi^{(2)i}$ just for   all the terms involving  $a_{n,m}$  and also interchange $D\phi^{(1)i}\leftrightarrow D\phi^{(2)i}$ for all $b_{n,m}$' terms and essentially add these terms to above couplings as well.

The other point should be clarified is that  here all scalar field's  covariant derivatives are indeed partial  derivatives. Notice that when fields change very slowly
then above couplings go back to two scalar, two tachyon couplings of tachyon DBI action. Although here we do not need to actually include gauge fields, however, it is worth trying to point out that

\beqa
D_{a_1}\cdots D_{a_n}T&=&\prt_{a_1}D_{a_2}\cdots D_{a_{n}}T-i(A^{(1)}_{a_1}-A^{(2)}_{a_1})D_{a_2}\cdots D_{a_{n}}T \nonumber\eeqa

It is highly recommended to mention once more that in this paper we discovered some new couplings. In particular likewise the first part of the amplitude to make sense of consistent result for brane anti brane system,
 all the couplings
 \beqa
D\phi^{(1)i}.D\phi^{(2)}_i\nonumber\eeqa
 and their new corrections must be appeared in brane anti brane effective action.

 \vskip .2in


Notice that these new corrections for brane anti brane systems \reef{hdts212} ( all order two tachyon two scalar couplings) can not be derived from the corrections of two tachyon and two gauge fields of the same system (compare equation (46) of \cite{Garousi:2008xp} with \reef{hdts212}). The reasons are as follows. First of all by direct computations we showed  that the sign of all  the coefficients of $b_{n,m}$ must get reversed and second of all while the coefficients of $L_1,L_2$ in both cases are the same , the coefficients of $L_3,L_4$ are completely different (which are not over all coefficients), that is why we claim the direct computations are needed. This clearly shows that by T-duality transformation, one can guess the general structure of higher derivative corrections but all their coefficients can just be fixed by direct Scattering computations as has been shown and argued in \cite{Hatefi:2012wj,Hatefi:2012zh,Hatefi:2012ve}.


 \vskip .2in

Now in order to show that we have obtained consistent higher derivative corrections of brane anti brane to all orders in $\alpha'$ , we would like to use these new couplings in \reef{hdts212} to indeed check out all infinite massless scalar poles of our amplitude $C\phi T_2 T_2$. This is a very important check to do so.

\vskip .1in

 Therefore consider one RR -$(p+1)$ and one scalar and two either $T_1$ or $T_2$ tachyons for brane anti brane system, the Feynman rule in (42) must be taken as well.


Here as we have mentioned earlier on (at the moment) we ignore some contact interactions and just focus on new higher derivative corrections of two scalars two tachyons of brane anti brane to be able to produce all the infinite scalar poles.

\vskip.2in

Now making use of the new two scalar two tachyon couplings of \reef{hdts212} and taking two possible permutations
,$ V_{\beta}^{j}(\phi,\phi_1,T_2,T_2)$  might be found as

\beqa V_{\beta}^{j}(\phi_1,\phi_1, T_2,T_2)&=&\xi_{1}^j  (-2i
T_p\pi)(\alpha')^{n+m+3}(a_{n,m}-b_{n,m}) \bigg(\frac{}{}(k_2\inn
k_1)^n(k\inn k_2)^m+(k_2\inn k_1)^n(k_3\inn k_1)^m
\nonumber\\&&+(k_2\inn k)^n(k_1\inn k_2)^m+(k\inn k_2)^n (k\inn
k_3)^m  +(k_3\inn k)^n(k_2\inn k)^m+(k_3\inn k)^n(k_3\inn
k_1)^m\nonumber\\&& +(k_1\inn k_3)^n(k_2\inn k_1)^m+(k_3\inn
k_1)^n(k_3\inn k)^m \bigg)\frac{1}{2}(s')(t')\Tr(\lam_1\lam_2\lam_3\lambda_{\beta}),\labell{verppttnn}\eeqa

Note that $\phi$ in the propagator of $G_{ij}(\phi)$ must be $\phi^{(1)},\phi^{(2)}$. Let us highlight that $T_2$ is related to the second component of complex scalar tachyon which we pointed out earlier. Considering   the following vertices

\beqa
G_{ij}(\phi) &=&\frac{i\delta_{ij} \delta_{\alpha\beta}}{(2\pi\alpha')^2 T_p
\left(u+t'+s'\right)}\nonumber\\
V_i(C_{p+1},\phi^{(1)})&=&i\mu_p(2\pi\alpha')\frac{1}{(p+1)!}\epsilon^{a_0\cdots a_{p}}H_{ia_0\cdots a_{p}}\Tr(\lambda_\alpha)\nonumber\\
V_i(C_{p+1},\phi^{(2)})&=&-i\mu_p(2\pi\alpha')\frac{1}{(p+1)!}\epsilon^{a_0\cdots a_{p}}H_{ia_0\cdots a_{p}}\Tr(\lambda_\alpha)\nonumber\\
V_{\beta}^{j}(\phi_2,\phi_1, T_2,T_2)&=&\xi_{1}^j (2i
T_p\pi)(\alpha')^{n+m+3}(a_{n,m}-b_{n,m}) \bigg(\frac{}{}(k_2\inn
k_1)^n(k\inn k_2)^m+(k_2\inn k_1)^n(k_3\inn k_1)^m
\nonumber\\&&+(k_2\inn k)^n(k_1\inn k_2)^m+(k\inn k_2)^n (k\inn
k_3)^m  +(k_3\inn k)^n(k_2\inn k)^m+(k_3\inn k)^n(k_3\inn
k_1)^m\nonumber\\&& +(k_1\inn k_3)^n(k_2\inn k_1)^m+(k_3\inn
k_1)^n(k_3\inn k)^m \bigg)\frac{1}{2}(s')(t')\Tr(\lam_1\lam_2\lam_3\lambda_{\beta}),\labell{verppttn22}\eeqa

and implementing  \reef{verppttn22} into (42) we get all infinite scalar poles in field theory as

  \beqa
&&8i\mu_p\frac{\eps^{a_{0}\cdots a_{p}}\xi_i H^{i}_{a_0\cdots
a_{p}}}{(p+1)!(s'+t'+u)}\Tr(\lam_1\lam_2\lam_3)
\sum_{n,m=0}^{\infty}(a_{n,m}-b_{n,m})[s'^{m}t'^{n}+s'^{n}t'^{m}]
  s't' \label{amphigh8}\eeqa

Simultaneously all the infinite scalar poles of the  string amplitude are

\beqa
{\cal A}^{\phi_1T_1T_1C}&\!\!\!\!=\!\!\!\!&\pm\frac{4i \mu_p}{(p+1 )!}\eps^{a_{0}\cdots a_{p}}H^i_{a_{0}\cdots a_{p}} \frac{\sum_{n,m=0}^{\infty}l_{n,m}(s'+t')^n(t's')^{m+1}}{(t'+s'+u)}  \xi_i
\labell{ww}\eeqa

\vskip.1in

Now if we have discovered \reef{hdts212} correctly, we would have to exactly produce all  the infinite scalar poles of string theory in field theory side as well.

Let us remove common factors and just compare string poles with field theory poles at each order of $\alpha'$.

 \vskip.1in

 By setting $n=m=0$, the string amplitude  gives rise $4 l_{0,0} s't'$ coefficient as it is shown in \reef{bvc}. At the same time at zeroth  order of $\alpha'$ the field theory amplitude gives us
\beqa
16s't'(a_{0,0}-b_{0,0})&=&-4\frac{\pi^2}{3} s't'\nonumber\eeqa
which is precisely equivalent to $4l_{0,0} s't'$ (the same coefficient in string side).

Now we can see that the new two scalar, two tachyon couplings of brane anti brane \reef{hdts212} work out.
In  $\alpha'$ order, from  string's amplitude we obtain
\beqa
4 l_{1,0} s't'(s'+t')\nonumber\eeqa

 and in particular in field 's amplitude we get to
\beqa
8 s't'(s'+t')(a_{1,0}+a_{0,1}-b_{0,1}-b_{0,1})&=&32\z(3)(s'+t') s't'=4l_{1,0} s't'(s'+t')\nonumber\eeqa
For $(\alpha')^2$ order, string amplitude related to
\beqa
4l_{2,0}s't'(s'+t')^2+4l_{0,1}(s't')^2  \nonumber\eeqa
 and field theory amplitude carries
\beqa
&&16 (s't')^2(a_{1,1}-b_{1,1}) +8s't'(a_{0,2}+a_{2,0}-b_{0,2}-b_{2,0})[(s')^2+(t')^2]\nonumber\\
&&=4s't'(-\frac{7\pi^4}{45}(s'+t')^2+\frac{\pi^4}{45}s't')=4(l_{2,0} (s'+t')^2+l_{0,1}s't')s't'\nonumber\eeqa
Finally we would like to check  $\alpha'^3$ order , that is, string amplitude  does carry
\beqa
4l_{3,0}(s'+t')^3t's'+4l_{1,1}(s'+t')(s't')^2  \nonumber\eeqa
 and field amplitude does include
\beqa
&&8s't'(a_{3,0}+a_{0,3}-b_{0,3}-b_{3,0})[(s')^3+(t')^3]+8s't'(a_{1,2}+a_{2,1}-b_{1,2}-b_{2,1})s't'(s'+t')\nonumber\\
&&=128\z(5)s't'(s'^3+t'^3)+8s'^2t'^2(s'+t')(32\z(5)+4\pi^2\z(3)/3)\nonumber\eeqa
which is exactly equivalent to the same coefficients $4l_{3,0}(s'+t')^3t's'+4l_{1,1}(s'+t')(s't')^2$ in string side.

 \vskip.2in

We can easily do check all the corrections up to all orders, so we reach to the important point that , we have exactly explored all order  two tachyon, two scalar couplings of brane anti brane systems.
 \vskip.2in

Remember  in order to get consistent result, some new coupling  like the multiplication of $D\phi^{(1)i}$ and $D\phi^{(2)}_i$, that is,
\beqa
D\phi^{(1)i}.D\phi^{(2)}_i\nonumber\eeqa

has to appear inside these infinite higher derivative corrections, this fact becomes obvious if we concentrate on  non zero $b_{n,m}$  coefficients in \reef{hdts212}.

\vskip.2in

Let us complete our discussions even for contact interactions which have been overlooked in producing infinite scalar poles for $p=n+2 $ case in field theory analysis. They are given by
\beqa
&&-i\mu_p(\alpha')^2\frac{ \eps^{a_{0}\cdots a_{p}}H^{i}_{a_{0}\cdots a_{p}}}{(p+1)!}
(-\xi_i(s')(t'))\sum_{n,m=0}^{\infty}(a_{n,m}-b_{n,m})\nonumber\\
&&\left[\left(2\sum_{\ell=1}^m \pmatrix{m\cr
\ell}(t'^{m-\ell}s'^n+s'^{m-\ell}t'^n)+2\sum_{\ell=1}^n \pmatrix{n\cr
\ell}(t'^{n-\ell}s'^m+s'^{n-\ell}t'^m)\right)(\alpha'k^2)^{\ell-1} \right.\nonumber\\
&&\left.+\sum_{\ell=1,j=1}^{n,m} \pmatrix{n\cr
\ell}\pmatrix{m\cr
j}(t'^{n-\ell}s'^{m-j}+s'^{n-\ell}t'^{m-j})(\alpha'k^2)^{\ell+j-1}\frac{}{}\right]\labell{masslesspole4}
\eeqa

 These couplings might be written down as
\beqa
&&i(\alpha')^2\mu_p\frac{ \eps^{a_{0}\cdots a_{p}}H^{i}_{a_{0}\cdots a_{p}}}{(p+1)!}
\left[-\xi_i s't'\right]\nonumber\\
&&\times \sum_{p,n,m=0}^{\infty}e'_{p,n,m}(s'+t'+u)^p(s'+t')^n(t's')^m\eeqa

\vskip .1in

Notice that  we can write $e'_{p,n,m}$ as $a_{n,m}$ and $b_{n,m}$. Some of the  contact interactions for $C \phi T_1 T_1$ may have precise structure like the above couplings so as a matter of fact one has to replace  the coefficients $e_{p,n,m}$ in  \reef{mmn} by $e_{p,n,m}-e'_{p,n,m}$

\vskip .2in

 Concerning some of the on-shell ambiguities, all we can express is that, one has to carry out a five point open super string computation like $TT\phi\phi A$ to over come some of the ambiguities however, we can not come over to this computation in this paper.
 This problem would remain  open question to the future research projects.

\vskip .2in

Eventually let us end up this section by producing all infinite contact interactions of the $C \phi T_1 T_1$ amplitude to all orders in $\alpha'$.  We just need to take into account the last term in \reef{mmn} such that , all contact interactions for this part of the amplitude can be summarized as

\beqa &&4i\mu_p\frac{ \eps^{a_{0}\cdots
a_{p}}H_{ia_{0}\cdots a_{p}}}{(p+1)!} \left[
\xi^i s't'\right]
\sum_{p,n,m=0}^{\infty}e_{p,n,m}(s'+t'+u)^p(s'+t')^n((t')(s'))^m\labell{contactterm}\eeqa

These terms can be reproduced in field theory by taking the following couplings
\beqa 2(\alpha')^2\mu_p\frac{ \eps^{a_{0}\cdots
a_{p}}H^{i}_{a_{0}\cdots a_{p}}}{(p+1)!}
\sum_{p,n,m=0}^{\infty}e_{p,n,m}(s'+t'+u)^p(s'+t')^n((t')(s'))^m\nonumber\\
\!\!\!\times
\left[\prt_b\prt_c(\phi_1-\phi_2)_iD^bT_1D^cT_1+T_1\rightarrow
T_2\right]\nonumber\eeqa

Note that in the above couplings the commutator in the definition of tachyon's covariant derivative should not  be considered as we do not have any external gauge field here.

In order to produce all the infinite contact interactions, the following derivative must act on the above coupling as well .

\beqa
((s')(t'))^mH\phi TT&\rightarrow&\left(\alpha'\right)^{2m}H\prt_{a_1}\cdots\prt_{a_{2m}}\phi D^{a_1}\cdots D^{a_m}TD^{a_{m+1}}\cdots D^{a_{2m}}T\nonumber\\
(s'+t')^nH \phi TT&\rightarrow&\left(\alpha'\right)^nH\prt^{a_1}\cdots \prt^{a_n}\phi D_{a_1}\cdots D_{a_n}(TT)\nonumber\\
(s'+t'+u)^pH \phi
TT&\rightarrow&\left(\frac{\alpha'}{2}\right)^pH(D_aD^a)^p(\phi
TT)\eeqa

One has to pay particular attention to the fact that the above couplings indeed have some on-shell ambiguities.

These ambiguities should be fixed  by computing the amplitude of $C TT \phi \phi$ in the world volume of brane anti brane systems. It is understood by field theory analysis that above couplings definitely will be appeared in the infinite tachyon poles of  $C TT \phi \phi$ amlitude. It would be interesting to carry out this computation  in detail.

\section{Conclusions}

In this paper using direct computations, we have obtained the closed form of the amplitude of one RR, two tachyons and one scalar field in the world volume of brane anti brane systems. We discovered that  the vertex of two tachyons and one gauge field $V^{a}(A  ,T_1,T_2)$ should not receive any correction and in particular by obtaining the infinite higher derivative corrections of one RR, one scalar and one gauge field of BPS branes, we could explore all the  infinite gauge poles of the amplitude of $C\phi TT$ for $p=n$ case. The special expansion (with taking into account the related Feynman rules) dictates us that leading terms do correspond to the DBI and Wess-Zumino effective actions, however all the other non leading terms in the expansion should be related to  their higher derivative corrections.

\vskip .2in
In the case of D-brane/anti D-brane the derivative expansion cannot
be thought of as a low energy expansion since the mass of the tachyon is of
order of the string scale. Nevertheless we argued that a particular
kinematic limit can be used to define a related expansion. It is very important to emphasize that we have discovered all the higher derivative corrections to two tachyon, two scalar field couplings in the world volume of brane anti brane systems to all orders in $\alpha'$ in \reef{hdts212} and check them out by producing all the infinite scalar   $(u+s'+t')$-channel poles of our amplitude for $p+2=n$ case.
\vskip .2in

To make sense of super string computations, one has to distinguish the new corrections of brane anti brane systems in \reef{hdts212} from non-BPS branes' corrections in \reef{hdts}.

\vskip .2in

 Infinite gauge poles have provided  remarkable information to actually derive
all infinite higher derivative corrections of $\partial_i C_{p-1}\wedge F(\phi_1+\phi_2)^i$.

By analysing  contact interactions of the amplitude for $p=n$ case we obtained a new coupling like $\partial_i C_{p-1} \wedge DT \wedge DT^*(\phi_1+\phi_2)^i $ and fixed its coefficient, we also found all its infinite extensions.

 \vskip .1in

 Note that, there is no correction to $\Tr(\phi^i)  H_{ia_0 \cdots a_p} \epsilon^{a_0 \cdots a_p}$, therefore all higher scalar poles have provided the complete information for infinite higher derivative corrections to two scalars, two tachyons of brane anti brane systems. Contact interactions for $p+2=n$ give us remarkable clues, not only on new coupling like $\epsilon^{a_0 \cdots a_p} H_{ia_0 \cdots a_p} (\phi^{(1)}-\phi^{(2)})^i TT^*$ but also
 on infinite higher derivative corrections thereof.  We also derived a new coupling like

\beqa 2(\alpha')^2\mu_p\frac{ \eps^{a_{0}\cdots
a_{p}}H^{i}_{a_{0}\cdots a_{p}}}{(p+1)!}
\sum_{p,n,m=0}^{\infty}e_{p,n,m}(s'+t'+u)^p(s'+t')^n((t')(s'))^m\nonumber\\
\!\!\!\times
\left[\prt_b\prt_c(\phi_1-\phi_2)_iD^bT_1D^cT_1+T_1\rightarrow
T_2\right]\nonumber\eeqa
 it would be interesting to see whether or not commutator in the definition of the covariant derivative of tachyon can be held. To settle this remark, one must carry out either $CTT \phi A$ or $CTT \phi AA$ amplitude in the world volume of brane anti brane systems. It is interesting to perform these amplitudes and find out all infinite higher derivative corrections to their effective actions as well. It would be interesting to check $CTTTT$ amplitude to eventually fix symmetrized trace tachyon effective action in the world volume of brane anti brane systems \cite{Hatefi:2013eh}.
Finally we point out that it might be interesting to
consider the thermodynamical aspects of such an action. It has been argued
that, at finite temperature such D-brane/ anti D-brane systems can be stable
and related to black holes.
\section*{Acknowledgments}

I would like to thank A.Sen, Joe.Polchinski, G.Thompson, F.Quevedo, Rob Myers, A.Belavin,
 N.Lambert, A.Lerda, A.Sagnotti, G.Veneziano, R.K Gupta and in particular K.S.Narain, J.Maldacena and L.Alvarez-Gaume for several valuable discussions.
I also thank the referee for providing several fascinating comments/future directions.


\begin{thebibliography}{2007}
\bibitem{Polchinski:1995mt}
  J.~Polchinski,
  ``Dirichlet Branes and Ramond-Ramond charges,''
  Phys.\ Rev.\ Lett.\  {\bf 75}, 4724 (1995)
  [hep-th/9510017].
\bibitem{Witten:1995im}
  E.~Witten,
  ``Bound states of strings and p-branes,''
  Nucl.\ Phys.\ B {\bf 460}, 335 (1996)
  [hep-th/9510135].
\bibitem{Polchinski:1996na}
  J.~Polchinski,
  ``Tasi lectures on D-branes,''
  hep-th/9611050.
\bibitem{Bachas:1998rg}
  C.~P.~Bachas,
  ``Lectures on D-branes,''
  [hep-th/9806199].
\bibitem{Hatefi:2012zh}
  E.~Hatefi,
   ``Shedding light on new Wess-Zumino couplings with their corrections to all orders in alpha-prime,''
  JHEP {\bf 1304}, 070 (2013)
  [arXiv:1211.2413 [hep-th]].




\bibitem{Hatefi:2010ik}
  E.~Hatefi,
  ``On effective actions of BPS branes and their higher derivative corrections,''
  JHEP {\bf 1005}, 080 (2010)
  [arXiv:1003.0314 [hep-th]].
\bibitem{Myers:1999ps}
  R.~C.~Myers,
  ``Dielectric branes,''
  JHEP {\bf 9912} (1999) 022
  [hep-th/9910053].
\bibitem{Li:1995pq}
  M.~Li,
  ``Boundary states of D-branes and Dy strings,''
  Nucl.\ Phys.\ B {\bf 460}, 351 (1996)
  [hep-th/9510161].
\bibitem{Douglas:1995bn}
  M.~R.~Douglas,
  ``Branes within branes,''
  In *Cargese 1997, Strings, branes and dualities* 267-275
  [hep-th/9512077].
\bibitem{Green:1996dd}
  M.~B.~Green, J.~A.~Harvey and G.~W.~Moore,
  ``I-brane inflow and anomalous couplings on d-branes,''
  Class.\ Quant.\ Grav.\  {\bf 14} (1997) 47
  [hep-th/9605033].

\bibitem{Gutperle:2002ai}
  M.~Gutperle and A.~Strominger,
   ``Space - like branes,''
  JHEP {\bf 0204}, 018 (2002)
  [hep-th/0202210].

\bibitem{Sen:2002nu}
  A.~Sen,
   ``Rolling tachyon,''
  JHEP {\bf 0204}, 048 (2002)
  [hep-th/0203211].

\bibitem{Sen:2002in}
  A.~Sen,
   ``Tachyon matter,''
  JHEP {\bf 0207}, 065 (2002)
  [hep-th/0203265].

\bibitem{Lambert:2003zr}
  N.~D.~Lambert, H.~Liu and J.~M.~Maldacena,
   ``Closed strings from decaying D-branes,''
  JHEP {\bf 0703}, 014 (2007)
  [hep-th/0303139].
\bibitem{Sen:2004nf}
  A.~Sen,
   ``Tachyon dynamics in open string theory,''
  Int.\ J.\ Mod.\ Phys.\ A {\bf 20}, 5513 (2005)
  [hep-th/0410103].









\bibitem{Sen:1999mg}
  A.~Sen,
  ``NonBPS states and Branes in string theory,''
  hep-th/9904207.
%

\bibitem{Garousi:2008ge}
  M.~R.~Garousi and E.~Hatefi,
 ``More on WZ action of non-BPS branes,''
  JHEP {\bf 0903}, 008 (2009)
  [arXiv:0812.4216 [hep-th]].
\bibitem{Hatefi:2012wj}
  E.~Hatefi,
  ``On higher derivative corrections to Wess-Zumino and Tachyonic actions in type II super string theory,''
  Phys.\ Rev.\ D {\bf 86}, 046003 (2012)
  [arXiv:1203.1329 [hep-th]].





\bibitem{Garousi:2007fk}
  M.~R.~Garousi and E.~Hatefi,
   ``On Wess-Zumino terms of Brane-Antibrane systems,''
  Nucl.\ Phys.\ B {\bf 800} (2008) 502
  [arXiv:0710.5875 [hep-th]].
\bibitem{Sen:2003tm}
  A.~Sen,
   ``Dirac-Born-Infeld action on the tachyon kink and vortex,''
  Phys.\ Rev.\ D {\bf 68}, 066008 (2003)
  [hep-th/0303057].


\bibitem{Sen:1998sm}
  A.~Sen,
  ``Tachyon condensation on the brane anti-brane system,''
  JHEP {\bf 9808}, 012 (1998)
  [hep-th/9805170].
\bibitem{Kennedy:1999nn}
  C.~Kennedy and A.~Wilkins,
    ``Ramond-Ramond couplings on Brane - anti-Brane systems,''
  Phys.\ Lett.\ B {\bf 464}, 206 (1999)
  [hep-th/9905195].
\bibitem{Sen:1998rg}
  A.~Sen,
   ``Stable nonBPS states in string theory,''
  JHEP {\bf 9806} (1998) 007
  [hep-th/9803194].

\bibitem{Bergman:1998xv}
  O.~Bergman and M.~R.~Gaberdiel,
   ``Stable nonBPS D particles,''
  Phys.\ Lett.\ B {\bf 441}, 133 (1998)
  [hep-th/9806155]
  ;
  A.~Sen,
  ``SO(32) spinors of type I and other solitons on brane - anti-brane pair,''
  JHEP {\bf 9809}, 023 (1998)
  [hep-th/9808141];
  M.~Frau, L.~Gallot, A.~Lerda and P.~Strigazzi,
  ``Stable nonBPS D-branes in type I string theory,''
  Nucl.\ Phys.\ B {\bf 564}, 60 (2000)
  [hep-th/9903123]
  ;
  E.~Dudas, J.~Mourad and A.~Sagnotti,
  ``Charged and uncharged D-branes in various string theories,''
  Nucl.\ Phys.\ B {\bf 620}, 109 (2002)
  [hep-th/0107081]
  ;
  E.~Eyras and S.~Panda,
   ``The Space-time life of a nonBPS D particle,''
  Nucl.\ Phys.\ B {\bf 584} (2000) 251
  [hep-th/0003033]
  ;
  E.~Eyras and S.~Panda,
  ``NonBPS branes in a type I orbifold,''
  JHEP {\bf 0105}, 056 (2001)
  [hep-th/0009224].

\bibitem{Witten:1998cd}
  E.~Witten,
  ``D-branes and K theory,''
  JHEP {\bf 9812}, 019 (1998)
  [hep-th/9810188].


\bibitem{Dvali:1998pa}
  G.~R.~Dvali and S.~H.~H.~Tye,
  ``Brane inflation,''
  Phys.\ Lett.\ B {\bf 450} (1999) 72
  [hep-ph/9812483].

\bibitem{Burgess:2001fx}
  C.~P.~Burgess, M.~Majumdar, D.~Nolte, F.~Quevedo, G.~Rajesh and R.~-J.~Zhang,
   ``The Inflationary brane anti-brane universe,''
  JHEP {\bf 0107}, 047 (2001)
  [hep-th/0105204].
\bibitem{Choudhury:2003vr}
  D.~Choudhury, D.~Ghoshal, D.~P.~Jatkar and S.~Panda,
   ``Hybrid inflation and brane - anti-brane system,''
  JCAP {\bf 0307}, 009 (2003)
  [hep-th/0305104].

\bibitem{Kachru:2003sx}
  S.~Kachru, R.~Kallosh, A.~D.~Linde, J.~M.~Maldacena, L.~P.~McAllister and S.~P.~Trivedi,
   ``Towards inflation in string theory,''
  JCAP {\bf 0310}, 013 (2003)
  [hep-th/0308055].

\bibitem{Casero:2007ae}
  R.~Casero, E.~Kiritsis and A.~Paredes,
   ``Chiral symmetry breaking as open string tachyon condensation,''
  Nucl.\ Phys.\ B {\bf 787}, 98 (2007)
  [hep-th/0702155 [HEP-TH]].
\bibitem{Dhar:2007bz}
  A.~Dhar and P.~Nag,
   ``Sakai-Sugimoto model, Tachyon Condensation and Chiral symmetry Breaking,''
  JHEP {\bf 0801}, 055 (2008)
  [arXiv:0708.3233 [hep-th]].


\bibitem{Kraus:2000nj}
  P.~Kraus and F.~Larsen,
  ``Boundary string field theory of the D anti-D system,''
  Phys.\ Rev.\ D {\bf 63}, 106004 (2001)
  [hep-th/0012198].

\bibitem{Roepstorff:1998vh}
  G.~Roepstorff,
  ``Superconnections and the Higgs field,''
  J.\ Math.\ Phys.\  {\bf 40}, 2698 (1999)
  [hep-th/9801040].

\bibitem{Hatefi:2012ve}
  E.~Hatefi and I.~Y.~Park,
  ``More on closed string induced higher derivative interactions on D-branes,''
  Phys.\ Rev.\ D {\bf 85} (2012) 125039
  [arXiv:1203.5553 [hep-th]].
;
  E.~Hatefi,
  ``Three Point Tree Level Amplitude in Superstring Theory,''
  Nucl.\ Phys.\ Proc.\ Suppl.\  {\bf 216} (2011) 234
  [arXiv:1102.5042 [hep-th]].
;
  E.~Hatefi,
  ``Closed string Ramond-Ramond proposed higher derivative interactions on fermionic amplitudes in IIB,''
  arXiv:1302.5024 [hep-th].



\bibitem{Hatefi:2012rx}
  E.~Hatefi and I.~Y.~Park,
  ``Universality in all-order $\alpha'$ corrections to BPS/non-BPS brane world volume theories,''
  Nucl.\ Phys.\ B {\bf 864} (2012) 640
  [arXiv:1205.5079 [hep-th]];
  E.~Hatefi,
  ``All order $\alpha'$ higher derivative corrections to non-BPS branes of type IIB Super string theory,''
  JHEP {\bf 1307}, 002 (2013)
  [arXiv:1304.3711 [hep-th]];
  E.~Hatefi,
  ``Selection rules, RR couplings on non-BPS branes and their all order $\alpha'$-corrections in type IIA(B) super string theories,''
  arXiv:1307.3520 [hep-th].

\bibitem{Hatefi:2012bp}
  E.~Hatefi, A.~J.~Nurmagambetov and I.~Y.~Park,
  ``ADM reduction of IIB on $\mathcal{H}^{p,q}$ to dS braneworld,''
  JHEP {\bf 1304}, 170 (2013)
  [arXiv:1210.3825 [hep-th]].


\bibitem{Hatefi:2012sy}
  E.~Hatefi, A.~J.~Nurmagambetov and I.~Y.~Park,
  ``$N^3$ entropy of $M5$ branes from dielectric effect,''
  Nucl.\ Phys.\ B {\bf 866}, 58 (2013)
  [arXiv:1204.2711 [hep-th]].
;
  E.~Hatefi, A.~J.~Nurmagambetov and I.~Y.~Park,
   ``Near-Extremal Black-Branes with n*3 Entropy Growth,''
  Int.\ J.\ Mod.\ Phys.\ A {\bf 27}, 1250182 (2012)
  [arXiv:1204.6303 [hep-th]].

\bibitem{deAlwis:2013gka}
  S.~de Alwis, R.~K.~Gupta, E.~Hatefi and F.~Quevedo,
  ``Stability, Tunneling and Flux Changing de Sitter Transitions in the Large Volume String Scenario,''
  arXiv:1308.1222 [hep-th].


\bibitem{Liu:2001qa}
  H.~Liu and J.~Michelson,
  ``-trek III: The search for Ramond-Ramond couplings,''
  Nucl.\ Phys.\  B {\bf 614}, 330 (2001)
  [arXiv:hep-th/0107172];
  D.~Friedan, E.~J.~Martinec and S.~H.~Shenker,
  ``Conformal Invariance, Supersymmetry and String Theory,''
  Nucl.\ Phys.\ B {\bf 271} (1986) 93;
  V.~A.~Kostelecky, O.~Lechtenfeld, W.~Lerche, S.~Samuel and S.~Watamura,
  ``Conformal Techniques, Bosonization and Tree Level String Amplitudes,''
  Nucl.\ Phys.\ B {\bf 288}, 173 (1987).

\bibitem{Fotopoulos:2001pt}
  A.~Fotopoulos,
   ``On (alpha')**2 corrections to the D-brane action for non-geodesic
  world-volume embeddings,''
  JHEP {\bf 0109}, 005 (2001)
  [arXiv:hep-th/0104146].



\bibitem{Hatefi:2013eh}
  E.~Hatefi ,
  "Symmetrized trace D-brane anti D-brane effective actions and all order $\alpha'$ corrections to one RR and four tachyon couplings",[work in progress]





\end{thebibliography}
\end{document}